 \documentclass[accepted]{uai2023} 

\usepackage[american]{babel}

\usepackage{natbib} 
\usepackage{mathtools} 
\usepackage{booktabs} 
\usepackage{tikz} 

\usepackage{amsfonts, amssymb}
\usepackage{algorithm}
\usepackage[noend]{algpseudocode}
\usepackage{caption}
\usepackage{subcaption}
\usepackage{multirow}
\usepackage{arydshln}

\usepackage{tikz-qtree}
\usetikzlibrary{trees}
\usetikzlibrary{automata,positioning}

\usepackage{amsthm}

\newtheorem{theorem}{Theorem}

\theoremstyle{remark}

\theoremstyle{definition}

\newcommand{\blue}{\textcolor{blue}}
\def\ci{\perp\!\!\!\perp}
\newcommand{\red}{\textcolor{red}}

\DeclareMathOperator{\pa}{pa} 

\title{On Testability and Goodness of Fit Tests in Missing Data Models}

\author[1]{Razieh~Nabi\vspace{-0.3cm}}
\author[2]{Rohit~Bhattacharya}
\affil[$\text{\textcolor{white}{1}}$]{%
	{\hspace{-1.1cm} razieh.nabi@emory.edu \hspace{3.55cm} rb17@williams.edu \vspace{0.25cm}}
}
\affil[1]{%
	Department of Biostatistics and Bioinformatics\\
	Emory University\\
	Atlanta, Georgia, USA
}
\affil[2]{%
	Department of Computer Science\\
	Williams College\\
	Williamstown, Massachusetts, USA
}

\begin{document}
\maketitle

\begin{abstract}
	Significant progress has been made in developing identification and estimation techniques for missing data problems where modeling assumptions can be described via a directed acyclic graph. The validity of results using such techniques rely on the assumptions encoded by the graph holding true; however, verification of these assumptions has not received sufficient attention in prior work. In this paper, we provide new insights on the testable implications of three broad classes of missing data graphical models, and design goodness-of-fit tests for them. The classes of models explored are: sequential missing-at-random and  missing-not-at-random models which can be used for modeling longitudinal studies with dropout/censoring, and a  no self-censoring model which can be applied to cross-sectional studies and surveys. 
\end{abstract}

\section{Introduction}
\label{sec:intro}

Missing data is a common issue in applied problems. To infer a parameter of interest under missingness, often a statistical model is posed that encodes a set of assumptions on the missingness mechanisms. These assumptions  are commonly divided into three main types: missing-completely-at-random (MCAR) where missingness does not have a cause and hence complete-case analysis is justifiable, missing-at-random (MAR) where all causes of missingness are assumed to be fully observed, and missing-not-at-random (MNAR) where causes of missingness are either only partially observed and/or fully unobserved \citep{little2019statistical}. 

MNAR models are perhaps the most common form of missingness in practice, and the most challenging since observations are systematically missing; yet such models are  underused due to the complexity of the identification and estimation procedures needed to recover parameters of interest as functions of observed data law. A recent line of research has proposed to use causal graphical models as a representation of the statistical models for missing data \citep{mohan2013missing, thoemmes2014cautious, shpitser2016consistent, saadati2019adjustment}. 
A causal graph not only encodes conditional independence relations between variables but also depicts the causal mechanisms responsible for missingness, making it a useful tool for interpretation of the underlying assumptions \citep{glymour2006using, daniel2012using, martel2013definition, scharfstein2021markov}. Further, just as in causal inference, graphical representations of missingness allow for the design of algorithms that automate certain steps of identification and estimation schemes; see \cite{mohan2021graphical, nabi2022causal} for detailed reviews. 

While advances in graphical models of missing data have yielded useful insights into identifying and estimating parameters of interest, the validity of any result relies on the substantive assumptions encoded by the graph holding true. In order to confirm testability of a restriction in missing data models, we have to examine its implications on the observed data distribution; this enables the design of empirical testing procedures from finite (but partially unobserved) samples. Unfortunately, we may not always be able to test all the encoded restrictions. The permutation model, proposed by \cite{robins97non-a}, is an example of a graphical MNAR model that is untestable. \cite{mohan2014testability} provided examples of other impediments for testability in graphical missing data models. For instance, in many cases, the assumption that no variable influences its own missingness (a.k.a. lack of self-censoring causes) is untestable.

Nonetheless, there are MAR and MNAR models that entail empirically testable restrictions \citep{mohan2014testability, tian2015missing, gain2018missing, tu2019causal}. The contributions of this paper in this regard are two-fold: (i) We expand on testable implications of missing data models that resemble ordinary conditional independencies in the underlying full law, but manifest as generalized a.k.a. Verma independencies in the observed law; (ii) We design empirical tests for restrictions in three broad classes of missing data models that use ideas from weighted likelihood-ratio tests and odds-ratio parameterizations of joint distributions. The model classes are:
\begin{enumerate}
	\item Sequential MAR models where missingness at each time step depends only on past observed values, 
	
	\item Sequential MNAR  models where missingness at each time step may depend on past observed variables as well as future unmeasured/missing values, and 
	
	\item MNAR models where missingness of each variable may depend on missing values of any other variable except itself.  
\end{enumerate} 
\vspace{-0.1cm}
%
The first two classes are particularly applicable for modeling missingness mechanisms in longitudinal studies with censoring, while the third class is more suitable for survey studies and situations where there is no inherent time ordering in the data collection process. With these insights, we can use partially observed data to gain information about the underlying missingness mechanisms and assess the adequacy of our chosen missingness models in our analyses. Additionally, we explore the extension of our results to scenarios where some variables are completely unmeasured or latent, further broadening the scope of our framework. Our results are also relevant for discovery and model selection tasks where the goal is not only to uncover the substantive relationships between the variables of interest but also to identify the processes that drive their missingness. \\
All proofs are deferred to the supplementary materials. 

\section{Notations and Preliminaries}
\label{sec:prelims}

Let $X = \{X_1, \ldots, X_K\}$ be a set of $K$ random variables with probability distribution $p(X).$ We denote the values of $X_k \in X$ by lower case letter $x_k \in \mathfrak{X}_k$, where $\mathfrak{X}_k$ denotes the state space of $X_k$. We assume a sample of $n$ i.i.d observations with missing values. To locate the missing cases, we consider a set of binary missingness indicators $R = \{R_1, \ldots, R_K\},$ where $R_k = 0$ when $x_k$ is missing and $R_k = 1$ when $x_k$ is observed. Let $X^* = \{X^*_1, \ldots, X^*_K\}$ denote the set of proxy random variables that represent the values of variables in $X$ that we actually observe. Each $X^*_k \in X^*$ is deterministically defined in terms of $R_k \in R$ and $X_k \in X$ as follows: if $R_k = 1, X^*_k = X_k,$ otherwise $X^*_k = ``?".$ We refer to $p(X)$ as the \emph{target law}, $p(R \mid X)$ as the \emph{missingness mechanism}, $p(X, R)$ as the \emph{full law}, and $p(R, X^*)$ as the \emph{observed data law.}   

A missing data model is a set of distributions defined over variables in  $\{X, R, X^*\}.$\footnote{For simplicity of notations, we assume all variables have missing values. All discussions however, can be easily generalized to scenarios where a subset of variables are fully observed.} Following the conventions in \cite{mohan2013missing}, we represent the missing data model via a directed acyclic graph (DAG) ${\cal G}(V)$, where vertices $V$ correspond to random variables in  $\{X, R, X^*\}$. In addition to acyclicity, a missing data DAG (or m-DAG\footnote{The term ``mDAG'' has also been used by \citep{evans2016graphs} to denote marginalized DAGs.} for short) imposes certain restrictions on the edges: variables in $R$ cannot point to variables in $X,$ and each $X^*_k \in X^*$ has only two parents: $X_k$ and $R_k$ (due to deterministic relations.) Similar to \cite{bhattacharya19mid}, we also allow for $X^*_i \rightarrow R_j$ edges. A few examples of m-DAGs are illustrated in Sections~~\ref{sec:testable-implications} and \ref{sec:tests}; edges corresponding to deterministic relations are drawn in gray. A full law  $p(X, R, X^*)$ that is Markov relative to ${\cal G}(V)$  factorizes as follows: 
\begin{align}
	\prod_{V_i \in X \cup R}  p(V_i \mid \pa_{\cal G}(V_i)) \times \prod_{X^*_k \in X^* } p(X^*_k \mid R_k, X_k),
	\label{eq:factor}
\end{align}%
where $\pa_{\cal G}(V_i)$ denotes the parents of $V_i$ in ${\cal G}(V).$ For convenience, we drop the  deterministic terms, $p(X^*_k | R_k, X_k)$, when discussing the factorization of the full law. 

The full law $p(X, R)$ is identified (can be expressed as a function of the observed data)  if and only if the missingness mechanism $p(R | X)$ is identified; the target law $p(X)$ is identified if and only if $p(R = 1 | X)$ is identified. Thus, identification of the full law implies that the  target law (and any function of the full law) is identified, but the reverse is not true. The missingness mechanism in an m-DAG factorizes as $\prod_{k} p(R_k | \pa_{\cal G}(R_k)),$ where the conditional density  $p(R_k | \pa_{\cal G}(R_k))$ is referred to as the \emph{propensity score} of $R_k. $ 

Numerous identification strategies in the field of graphical missing data literature focus on identifying each propensity score in a specific order, whether it is a total or partial order; notable works in this area include \cite{shpitser2015missing} and \cite{bhattacharya19mid}. In essence, to identify the propensity score of $R_k$, one can verify if $R_k$, given parents, is independent of the corresponding missingness indicators of its parents that are counterfactuals. If this condition holds, the propensity score can be identified using a simple argument based on conditional independence (d-separation). However, if this condition is not satisfied, one needs to examine whether it holds in post-fixing distributions, which are obtained through the recursive application of the \textit{fixing} operator. This operation involves inverse weighting the current distribution by the propensity score of the variable being fixed \citep{bhattacharya2022semiparametric, richardson2023nested}. We employ similar strategies in this work.

Similar to regular DAGs, absence of an edge in a missing data DAG ${\cal G}(V)$ entails conditional independence restrictions between the endpoint variables in the underlying distribution $p(V).$ These restrictions can be directly read off from the graph using Markov properties and d-separation rules  \citep{pearl2009causality} -- given disjoint sets $U, W, Z \subset V,$ the global Markov property states that if $U \ci_\text{d-sep} W \mid Z$ in ${\cal G}(V)$, then $U \ci W \mid Z$ in $p(V).$ 
In this work, we focus on restrictions where all variables are at least partially observed, which allows us to narrow our focus to testability of ordinary conditional independence restrictions in the full law. However, as we will see, even ordinary restrictions in the full law may manifest as generalized equality restrictions in the observed law. Testability of generalized equality restrictions induced by latent variables in m-DAGs is a challenging problem left for future work; see \cite{tian2002testable, shpitser2006identification, bhattacharya2022semiparametric} for more details on such restrictions when there is no missingness.

Unlike regular DAGs where the independence constraints can be tested using observed samples from the joint distribution, a conditional independence restriction in an m-DAG might be empirically untestable, or may manifest as more complex restrictions on the observed data law. If all the restrictions encoded in an m-DAG are provably untestable (i.e., no restriction on the observed data law), the full law Markov relative to the m-DAG is said to be \emph{nonparametric saturated} (as defined by \cite{robins97non-a}). Nonetheless, submodels of saturated missing data models may still be testable. In this paper, we discuss testability of assumptions in the three aforementioned classes of missing data models as submodels of two known saturated models: the \textit{permutation} model and the \textit{no self-censoring} model. 

\cite{robins97non-a} introduced the permutation model as follows: given an ordering on variables in $X,$ indexed by $k \in \{1, \ldots, K\}$, each missingness indicator $R_k$ is independent of the current and past variables in $X$ given the past observed variables in $R, X^*$ and future variables in $X.$ Formally, the model is defined via the following set of conditional independence restrictions:  
\begin{align}
	R_k \ci X_{\prec k+1} \mid R_{\prec k}, X^*_{\prec k}, X_{\succ k}, \ \forall k \quad \textit{\small (permutation)}
	\label{eq:perm}
\end{align}%
where $V_{\prec k} = \{V_1, \ldots, V_{k - 1}\}, V_{\succ k} = \{V_{k+1}, \ldots, V_K\}.$ \cite{robins97non-a} showed that the full law in this model is identified and is nonparametrically saturated. An m-DAG representation of the permutation model with $K=2$ variables is shown in Fig.~\ref{fig:seq-mar}(b). For a discussion on the substantive distinctions between $X^*_1 \rightarrow R_2$ and $X_1 \rightarrow R_2$ edges, refer to Appendix~A.3.

The no self-censoring model was introduced by \cite{shpitser2016consistent, sadinle2016itemwise}.\footnote{In \cite{sadinle2016itemwise}, the model is referred to as \textit{itemwise conditionally independent nonresponse} model.} The central assumption in this model is that no variable directly causes its own missingness status. Formally, the model is defined by the following set of conditional independence restrictions: 
\begin{align}
	R_k \ci X_k \mid R_{-k}, X_{-k}, \ \forall k \quad \textit{ (no self-censoring)}
	\label{eq:no-self}
\end{align}%
where $V_{-k} = V \setminus V_k.$ \cite{malinsky2021semiparametric} showed that this model is nonparametrically saturated and identified via an odds-ratio parameterization of the missingness mechanism; a description of this parameterization, which was proposed by \cite{chen2007semiparametric}, is provided in Appendix~A.1. The graphical representation of this model relies on a generalization of m-DAGs to allow for undirected edges between all pairs of $R$ vertices -- a graph with both directed and undirected edges is called a chain graph \citep{lauritzen1996graphical, shpitser2016consistent}.  An example of this model with $K=2$ variables is shown in Fig.~\ref{fig:seq-mnar_2}(b). The assumptions of the no self-censoring model are encoded in this chain graph by the following local Markov property: each missingness indicator $R_i$ is independent of all other variables on the graph given its neighboring missingness indicators (joined via an undirected edge $R_i - R_j$) and its parents ($X_j \rightarrow R_i$).

We now explore how one can select an appropriate m-DAG representation based on intuitive explanations of the missing data generation process. To illustrate this, imagine an investigator who is analyzing a large observational database that contains information on smoking habits and diagnostic test results for bronchitis among individuals in a city. In this scenario, let's consider $X_1$ as the true smoking status of an individual and $X_2$ as their bronchitis diagnosis. However, the investigator notices that there are missing entries in the database, which are indicated by the missingness indicators $R_1$ and $R_2$.
If we choose the permutation model depicted in Figure~\ref{fig:seq-mar}(b) to explain the missingness mechanism, it implies the following two processes: (i) $R_1 \leftarrow X_2$ suggesting that the measurement of an individual's smoking status depends on the counterfactual value of their bronchitis status; this may occur for example when a patient's smoking status is inquired on a suspected diagnosis of bronchitis before administering the test., and (ii)  $R_1 \rightarrow R_2 \leftarrow X^*_1$ suggesting that whether the true bronchitis status is measured via a diagnostic test depends on the doctor's awareness of the individual’s smoking status ($R_1$) and their observed value of smoking ($X_1^*$). 
On the other hand, the no self-censoring model shown in Fig.~\ref{fig:seq-mnar_2}(b) explains the missingness mechanism via the following three processes: (i) $R_1 \leftarrow X_2$ suggesting that a suspected diagnosis of bronchitis is likely to lead to an inquiry about the smoking status of the patient, (ii) $R_2 \leftarrow X_1$ suggesting that smokers are more likely to get tested for bronchitis, and (iii) $R_1 - R_2$ suggesting that ordering a diagnostic test for bronchitis increases the likelihood of ordering a test for bronchitis, and vice versa.

\section{New Insights into Testable Implications}
\label{sec:testable-implications}

Although the restrictions we study in this paper can be phrased in terms of ordinary independence restrictions in the full law of a missing data DAG model, they may only manifest in the observed data law via relatively complex functionals. In this section, we show that a d-separation statement between missingness indicators and substantive variables may correspond to generalized equality constraints, a.k.a. Verma constraints \citep{verma1990equivalence}, in the observed data distribution. This observation extends the current state of the art on testability in missing data models. 

\begin{figure}[!t] 
	\begin{center}
		\scalebox{0.68}{
			\begin{tikzpicture}[>=stealth, node distance=1.4cm]
				\tikzstyle{format} = [thick, circle, minimum size=1.0mm, inner sep=0pt]
				\tikzstyle{square} = [draw, thick, minimum size=4.5mm, inner sep=3pt]
				
				\begin{scope}[xshift=0cm]
					\path[->, thick]
					node[format] (x11) {$X_1$}
					node[format, right of=x11, xshift=0.45cm] (x21) {$X_2$}
					node[format, below of=x11] (r1) {$R_1$}
					node[format, below of=x21] (r2) {$R_2$}
					node[format, below of=r1] (x1) {$X^*_1$}
					node[format, below of=r2] (x2) {$X^*_2$}
					
					(x11) edge[blue] (x21) 
					(r1) edge[blue] (r2)
					(x1) edge[blue] (r2)
					
					(x11) edge[gray, bend right=25] (x1)
					(x21) edge[gray, bend left=25] (x2)
					(r1) edge[gray] (x1)
					(r2) edge[gray] (x2)
					
					node[format, below of=x1, xshift=1cm, yshift=0.75cm] (a) {(a)} ;
				\end{scope}
				
				\begin{scope}[xshift=3.8cm, yshift=0cm]
					\path[->, thick]
					node[format] (x11) {$X_1$}
					node[format, right of=x11, xshift=0.45cm] (x21) {$X_2$}
					node[format, below of=x11] (r1) {$R_1$}
					node[format, below of=x21] (r2) {$R_2$}
					node[format, below of=r1] (x1) {$X^*_1$}
					node[format, below of=r2] (x2) {$X^*_2$}
					
					(x11) edge[blue] (x21) 
					(x21) edge[blue] (r1)
					(r1) edge[blue] (r2)
					(x1) edge[blue] (r2)
					
					(x11) edge[gray, bend right=25] (x1)
					(x21) edge[gray, bend left=25] (x2)
					(r1) edge[gray] (x1)
					(r2) edge[gray] (x2)
					
					node[format, below of=x1, xshift=1cm, yshift=0.75cm] (b) {(b)} ;
				\end{scope}
				
				\begin{scope}[xshift=7.6cm, yshift=0cm]
					\path[->, thick]
					node[format] (x11) {$X_1$}
					node[format, right of=x11, xshift=0.45cm] (x21) {$X_2 = X^*_2$}
					node[format, below of=x11] (r1) {$R_1$}
					node[square, below of=x21] (r2) {$R_2=1$}
					node[format, below of=r1] (x1) {$X^*_1$}
					node[format, below of=r2] (x2) {}
					
					(x11) edge[blue] (x21) 
					(x21) edge[blue, dashed] (r1)
					
					(x11) edge[gray, bend right=25] (x1)
					(r1) edge[gray] (x1)
					
					node[format, below of=x1, xshift=1cm, yshift=0.75cm] (c) {(c)} ;
				\end{scope}
				
			\end{tikzpicture}
		}
		\vspace{-0.2cm}
		\caption{(a) Example of a MAR model; (b) Example of a  saturated permutation model; (c) The absence of $X_2 \rightarrow R_1$ edge in (a) can be tested in the intervention distribution $p(V \setminus R_2 | \text{do}(R_2=1))$ where the dashed edge indicates whether $p(V)$ is Markov wrt the MAR model in (a) or the permutation supermodel in (b). } 
		\label{fig:seq-mar} 
		\vspace{-0.6cm}
	\end{center}
\end{figure}
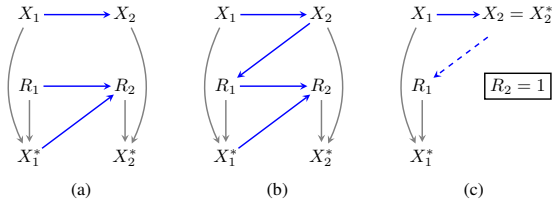 

Consider the m-DAG shown in Fig.~\ref{fig:seq-mar}(a): a MAR submodel of the permutation model in Fig.~\ref{fig:seq-mar}(b) where  $X_2 \rightarrow R_1$ is removed. Though the permutation model itself is nonparametric saturated, it is natural to ask if this MAR submodel, which encodes an additional d-separation relation $R_1 \ci X_2$, has a testable restriction on the observed data. 
To determine testability, we initially apply the criterion proposed by \cite{mohan2014testability}. According to their criterion, a d-separation condition displayed in an m-DAG ${\cal G}$ is testable if the missingness indicators associated with all partially observed variables involved in the relation are either already present in the separating set, or can be added to the set without spoiling the separation. By applying this criterion, we note that the  relation $R_1 \ci_\text{d-sep} X_2 | R_2$ does \emph{not} hold in ${\cal G}$ due to the open collider $R_2$ on the path $R_1 \rightarrow R_2 \leftarrow X^*_1 \leftarrow X_1 \rightarrow X_2.$ Therefore, one might conclude that $R_1 \ci X_2$ is not testable.  

Let us momentarily assume that $X$ in Fig.~\ref{fig:seq-mar}(a) consists of binary variables. We can compare number of parameters in the full law using (\ref{eq:factor}) against the saturated observed data law using pattern-mixture factorization \cite{rubin76inference} given by the marginal distribution of $R$ and conditional distribution of $X^*$ given $R$. The full law in Fig.~\ref{fig:seq-mar}(a) requires $7$ parameters ($3$ for $p(X)$, $1$ for $p(R_1)$, and $3$ for $p(R_2 | R_1, X^*_1)$) which is less than the number of parameters in the saturated observed law which is $8$ ($3$ for $p(R)$ and $5$ for $p(X^* | R)$.) Hence, it can be inferred that the additional restriction $R_1 \ci X_2$ (implied by the absence of $X_2 \rightarrow R_1$ edge) imposes constraints on the observed data law, at least in the discrete case.\footnote{Appendix~A.2 contains the parameter counting arguments.}   Interestingly, this contradicts the conclusion from the previous paragraph stating that $R_1 \ci X_2$ is not testable. Such contradictions are expected, however, as Mohan and Pearl's criterion is sufficient but not necessary for testability. 
 
Now, we aim to demonstrate that $R_1 \ci X_2$ is indeed testable, not only in discrete cases but also in more general nonparametric settings. In Fig.~\ref{fig:seq-mar}(a), conditioning on $R_2$ opened up the collider $R_1 \rightarrow R_2 \leftarrow X^*_1$ on the path from $R_1$ to $X_2$. From a causal perspective, removal of these edges corresponds to an intervention in which $R_2$ is set to a specific value.\footnote{We borrow the notion of intervention from causal inference by viewing each missingness indicator as a ``treatment variable''; see \citep{nabi2022causal} for details.} This results in the m-\emph{conditional} DAG (m-CDAG), as shown in Fig.~\ref{fig:seq-mar}(c). Due to  determinism (alternatively consistency), once $R_2$ is set to $1$, then $X_2 = X^*_2$ forming a single (observed) node. Following standard notation in \cite{pearl2009causality} we denote the intervention where we set $R_2$ to 1 as $\text{do}(R_2=1)$ and the corresponding intervention distribution as $p(X, R \setminus R_2, X^* | \text{do}(R_2=1)),$ or $p(. | \text{do}(R_2=1))$ for short. This intervention distribution can be obtained via truncation of the full law factorization where the propensity score of $R_2$, $p(R_2 | \pa_{\cal G}(R_2))$, is dropped. That is, $p(. | \text{do}(R_2=1)) = p(X, R, X^*) / p(R_2 | R_1, X^*_1)\vert_{R_2=1}$ and it factorizes according to the m-CDAG shown in Fig.~\ref{alg:seq-mar}(c). The relation $R_1 \ci_\text{d-sep} X_2 $ holds in the resulting m-CDAG, and $X_2$ is now fully observed. Further, the  propensity score of $R_2$  that takes us to the intervention distribution is a function of observed data. These facts combined imply that $R_1 \ci X_2$ imposes a restriction on the observed data in the form of a Verma constraint; i.e., a d-separation statement in an identified intervention distribution. 

The above example illustrates the core idea  for essential extensions of the previous testability criterion  \citep{mohan2014testability, mohan2021graphical}. We state that a d-separation condition displayed in an m-DAG is also testable if the missingness indicators associated with all partially observed variables involved in the relation can be intervened on (or, in other words, their corresponding propensity scores are identified) without spoiling the separation. We formalize this extension of testable restrictions in the next section (and partly in the appendix), where we also consider testability of independence statements between proxy variables and missingness indicators, and among missingness indicators themselves. 

\section{Testable Implications and Goodness-of-Fit Tests}
\label{sec:tests}

In this section, we investigate independence assumptions in the full law and their implications on the observed data law in three broad classes of missing data models, and provide ways of empirically evaluating these constraints. We formulate the testability criteria and goodness-of-fit tests for the general case of a missing data model with $K$ variables and illustrate the steps via examples. We consider likelihood-ratio tests for evaluating the independence $A \ci B \mid C$, which is typically performed by fitting $p(A \mid C)$ and $p(A \mid B, C)$ and comparing their goodness-of-fit. Under the null hypothesis of independence, both models should fit the data equally well. In addition to likelihood-ratio tests, we consider evaluating $A \ci B \mid C$ by computing the odds ratio of $A$ and $B$ conditioned on $C$ -- the independence relation holds \textit{if and only if} the odds ratio equals one for all values of $A, B, C$. Therefore, if under the alternative hypothesis of dependence, the odds ratio still equals one (with statistical significance-level $\alpha$), then the data agrees with the independence relation. In the following discussion, we let $\mathcal{M}_o$ denote the statistical model where the independence relation holds (the \textit{null} hypothesis), and let $\mathcal{M}_a$ denote the statistical supermodel where the independence relation does not hold (the \textit{alternative} hypothesis). 

\subsection{Sequential MAR models}

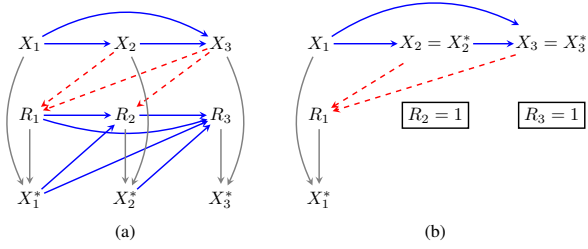
\begin{figure}[!t] 
	\begin{center}
		\scalebox{0.68}{
			\begin{tikzpicture}[>=stealth, node distance=1.4cm]
				\tikzstyle{format} = [thick, circle, minimum size=1.0mm, inner sep=0pt]
				\tikzstyle{square} = [draw, thick, minimum size=4.5mm, inner sep=3pt]
				
				\begin{scope}[xshift=0cm]
					\path[->, thick]
					node[format] (x11) {$X_1$}
					node[format, right of=x11, xshift=0.45cm] (x21) {$X_2$}
					node[format, right of=x21, xshift=0.45cm] (x31) {$X_3$}
					node[format, below of=x11] (r1) {$R_1$}
					node[format, below of=x21] (r2) {$R_2$}
					node[format, below of=x31] (r3) {$R_3$}
					node[format, below of=r1, yshift=-0.25cm] (x1) {$X^*_1$}
					node[format, below of=r2, yshift=-0.25cm] (x2) {$X^*_2$}
					node[format, below of=r3, yshift=-0.25cm] (x3) {$X^*_3$}
					
					(x11) edge[blue] (x21) 
					(r1) edge[blue] (r2)
					(x1) edge[blue] (r2)
					
					(x31) edge[red, dashed] (r2)
					(x31) edge[red, dashed] (r1)
					(x21) edge[red, dashed] (r1)
					
					(x21) edge[blue] (x31) 
					(r2) edge[blue] (r3)
					(x2) edge[blue] (r3)
					
					(x11) edge[blue, bend left] (x31) 
					(r1) edge[blue, bend right=17] (r3)
					(x1) edge[blue] (r3)
					
					(x11) edge[gray, bend right=25] (x1)
					(x21) edge[gray, bend left=25] (x2)
					(x31) edge[gray, bend left=25] (x3)
					(r1) edge[gray] (x1)
					(r2) edge[gray] (x2)
					(r3) edge[gray] (x3)
					
					node[format, below of=x2, xshift=0cm, yshift=0.75cm] (a) {(a)}; 
				\end{scope}
				
				\begin{scope}[xshift=5.6cm]
					\path[->, thick]
					node[format] (x11) {$X_1$}
					node[format, right of=x11, xshift=0.85cm] (x21) {$X_2 = X^*_2$}
					node[format, right of=x21, xshift=0.85cm] (x31) {$X_3=X^*_3$}
					node[format, below of=x11] (r1) {$R_1$}
					node[square, below of=x21] (r2) {$R_2=1$}
					node[square, below of=x31] (r3) {$R_3=1$}
					node[format, below of=r1, yshift=-0.25cm] (x1) {$X^*_1$}
					node[format, below of=r2, yshift=-0.25cm] (x2) {}
					node[format, below of=r3, yshift=-0.25cm] (x3) {}
					
					(x11) edge[blue] (x21) 
					
					(x31) edge[red, dashed] (r1)
					(x21) edge[red, dashed] (r1)
					
					(x21) edge[blue] (x31) 
					
					(x11) edge[blue, bend left] (x31) 
					
					(x11) edge[gray, bend right=25] (x1)
					(r1) edge[gray] (x1)
					
					node[format, below of=x2, xshift=0cm, yshift=0.75cm] (b) {(b)}; 
				\end{scope}
			\end{tikzpicture}
		}
		\vspace{-0.2cm}
		\caption{ (a) Example of a sequential MAR model (without the dashed edges) along with its permutation supermodel (with the dashed edges); (b) The graph Markov wrt the intervention distribution $p(. | \text{do}(R_2=1, R_3=1))$.} 
		\label{fig:seq-mar2}
		\vspace{-0.6cm}
	\end{center}
\end{figure}

We call a missing data model a \textit{sequential MAR} model if under an ordering $\prec$ that indexes variables by $k = 1, \ldots, K$, the following set of independence restrictions hold:
\begin{align}
	R_k \ci X \mid R_{\prec k}, X^*_{\prec k}, \ \forall k  \qquad \textit{(sequential-MAR)}
	\label{eq:seq-mar}
\end{align}%
Examples of this model are shown in Fig.~\ref{fig:seq-mar}(a) and Fig.~\ref{fig:seq-mar2}(a) (without the dashed edges). In addition to restrictions of a permutation model described in (\ref{eq:perm}), the sequential MAR model assumes $R_k \ci X_{\succ k} |  R_{\prec k}, X^*_{\prec k}, \ \forall k$ (it is straightforward to see this using graphoid axioms; see e.g., \citep{lauritzen1996graphical} for description of the axioms). Thus, we can view the sequential MAR model as a submodel of the permutation model. Since assumptions imposed by the permutation model alone are untestable, we focus on testable implications of these extra assumptions and propose ways to empirically evaluate them. 

The independence  $R_k \ci X_{\succ k} | R_{\prec k}, X^*_{\prec k}$  would be easily testable using observed data  if we could add $R_{\succ k} = 1$ in the conditioning set and thus evaluate the restriction using only observed cases of $X_{\succ k}$. Unfortunately, the independence  no longer holds if we condition on $R_{\succ k}$ (this is easily confirmed from discussion in the previous section and Fig.~\ref{fig:seq-mar2}(a).) However, we can instead intervene on $R_{\succ k}$ and check if the independence holds in the intervention distribution. The following theorem formalizes that restrictions in sequential MAR models defined above can always be tested as Verma constraints, i.e., (i) the independence holds in the corresponding m-CDAG, and (ii) the required intervention distributions are identified from observed data. 
\begin{theorem} \label{thm:seq-mar} 
	The independence $R_k \ci X_{\succ k} | R_{\prec k}, X^*_{\prec k}$ has a testable implication on the observed data distribution in the form of a Verma constraint: $R_k \ci  X_{\succ k} | R_{\prec k}, X^*_{\prec k}, \text{do}(R_{\succ k}  = 1)$, where the intervention distribution $p(X, R \setminus R_{\succ k}, X^* | \text{do}(R_{\succ k} =1))$ is identified. 	
\end{theorem}
The intuition for this result will become clear as we  discuss testing such constraints using $n$ (finite) i.i.d samples (denoted by ${\cal D}_n$). 
One possibility is to use a likelihood-ratio test and compare goodness-of-fits between $p(R_k | R_{\prec k}, X^*_{\prec k})$ and $p(R_k | R_{\prec k}, X^*_{\prec k}, X_{\succ k})$ \textbf{but} with respect to a distribution where $R_{\succ k}$ are intervened on and set to $1$. This intervention distribution is a truncated factorization of the full law where propensity scores of $R_{\succ k}$ are dropped, i.e., 

\vspace{-0.5cm}
{\small
	\begin{align*}
		p( . \mid \text{do}(R_{\succ k} = 1)) = \frac{p(V)}{\prod_{j \succ k} \ p(R_j \mid \pa_{\cal G}(R_j))}\Big|_{R_{\succ k} = 1}. 
	\end{align*}
}%
Let $W_{k}(\beta^o_k) \coloneqq p(R_k | R_{\prec k}, X^*_{\prec k}; \beta^o_k)$ and  $W_{k}(\beta^a_k) \coloneqq p(R_k | R_{\prec k}, X^*_{\prec k}, X_{\succ k}; \beta^a_k)$ (the null and alternative respectively.)
Estimating $\beta^o_k$ is relatively straightforward as $W_{k}(\beta^o_k)$ is a direct function of observed data, but estimating $\beta^a_k$ is more involved. We propose to estimate $\beta^a_k$, wrt the truncated/weighted distribution above. This entails using a weighted estimating equation where propensity scores of $R_{\succ k}$ are used as inverse weights to fit $\beta^a_k$. It is important to note however, that a propensity score $p(R_j \mid R_{\prec j}, X^*_{\prec j}, X_{\succ j})$ for any $R_j \in R_{\succ k},$  may itself need to be fitted via a weighted estimating equation, since $X_{\succ j}$ appears in the conditioning set and $R_j \not\ci R_{\succ j} | R_{\prec j}, X^*_{\prec j}, X_{\succ j}$. 

As an example, consider the sequential MAR model in Fig.~\ref{fig:seq-mar2}(a) (without the dashed edges). The null hypothesis $\mathcal{M}_o$ is the statistical model of this m-DAG and the alternative $\mathcal{M}_a$ is the permutation supermodel with the dashed edges.  We are interested in evaluating the independencies $R_1 \ci X_2, X_3$ and $R_2 \ci X_3 | R_1, X^*_1,$ which given Theorem~\ref{thm:seq-mar} translates into independence restrictions in $p(. | \text{do}(R_2=1, R_3=1))$ and $p(. | \text{do}(R_3=1))$, respectively. Testing $R_1 \ci X_2, X_3$ entails fitting $W_{r_1}(\beta^a_{r_1}) \coloneqq p(R_1 | X_2, X_3; \beta^a_{r_1})$ wrt the truncated/weighted factorization Markov relative to Fig.~\ref{fig:seq-mar2}(b) where $R_2$ and $R_3$ are intervened on and set to $1$. Thus, we can use propensity scores of $R_2$ and $R_3$ as inverse weights to estimate $\beta^a_{r_1}$. Let $\mathbb{P}_n[U(\beta^a_{r_1})] = 0$ be an unbiased estimating equation for $\beta^a_{r_1}$ wrt the full law ($\mathbb{P}_n[.] = \frac{1}{n} \sum_{i = 1}^n (.)$).  In other words, $\mathbb{P}_n[U(\beta^a_{r_1})]$ is any estimating equation that is unbiased for $\beta^a_{r_1}$ had there been no missingness. The following \textit{weighted} estimating equation then yields an unbiased estimator for $\beta^a_{r_1}$ wrt the observed data law:

\vspace{-0.55cm}
{\small
	\begin{align*}
		\mathbb{P}_n \bigg[  \frac{R_2 \times R_3}{p(R_2  \mid \pa_{\cal G}(R_2)) \times p(R_3 \mid \pa_{\cal G}(R_3))} \times U(\beta^a_{r_1}) \bigg] = 0, 
	\end{align*}
}%
where propensity score of $R_3$, $p(R_3 | R_1, R_2, X^*_1, X^*_2)$, can be fit using just observed data, denote it with $W_{r_3}(\widehat{\beta}_{r_3})$. However, fitting the propensity score of $R_2$, $p(R_2 | R_1, X^*_1, X_3),$ requires an intermediate step involving the intervention distribution where $R_3$ is intervened on and set to $1$, i.e.,  $p(X, R, X^*)/p(R_3 | \pa_{\cal G}(R_3))$ evaluated at $R_3=1$. Similar to the above logic, this entails a weighted estimating equation using $W_{r_3}(\widehat{\beta}_{r_3})$  as inverse weights to fit the propensity score of $R_2$, denoted by $W_{r_2}(\widehat{\beta}_{r_2})$. Now that we have a way of estimating $\beta^a_{r_1}$, we can test $R_1 \ci X_2, X_3$ using a weighted likelihood-ratio by computing 

\vspace{-0.55cm}
{\small
	\begin{align*}
		\rho = n\mathbb{P}_n\bigg[ \frac{R_2 \times R_3}{W_{r_2}(\widehat{\beta}_{r_2}) \times W_{r_3}(\widehat{\beta}_{r_3})} \times \log\Big( \frac{ W_{r_1}(\widehat{\beta}^a_{r_1})}{W_{r_1}(\widehat{\beta}^o_{r_1})} \Big) \bigg],
	\end{align*}
}%
where $W_{r_1}({\beta}^o_{r_1}) \coloneqq p(R_1; \beta^o_{r_1})$ and $\beta^o_{r_1}$ is simply the proportion of complete cases of $X_1,$ we can use likelihood chi-square or Wald tests to compare goodness-of-fits \citep{robins1997estimation, agostinelli2001test}. 

If we start our tests for the sequential MAR model by testing the restriction $R_2 \ci X_3 | R_1, X^*_1$, there are two possibilities: (i) The null might be rejected which immediately implies that the missing data model is not sequential MAR; (ii) The null is accepted which means $R_2$ does not have $X_3$ as a cause. In future tests, say for $R_1 \ci X_2, X_3$ in this case, this justifies fitting a simplified propensity score $p(R_2 | R_1,X^*_1, X^*_3)$ that makes use of all the observed data. This simplified propensity score also corresponds to the same model we would have already fit for the previous null hypothesis $\beta_{r_2}^o.$   This example reveals that there is a natural way to order the tests. For a model with $K$ variables, we would proceed backwards by first testing restrictions involving $R_{K-1}$, moving to $R_{K-2},$ and so on. If the current test succeeds, the corresponding model for the null can be re-used to produce weights for future estimating equations; if the test fails, then the assumptions of sequential MAR does not hold. Following such a sequence may help improve the power of each test by using all of the observed samples to estimate the weights in each step. We formalize this sequence of goodness-of-fit tests based on weighted likelihood-ratios in Algorithm~\ref{alg:seq-mar}, which takes an ordering  $\prec$ on the missingness indicators, null and alternative models as a tuple ${\cal M}$, and data samples  ${\cal D}_n$ as input. The $k^\text{th}$ iteration of the for loop concerns testing the independence $R_k \ci X_{\succ k} \mid R_{\prec k}, X^*_{\prec k}$. Note however, that as we proceed with the tests, we are restricted to fewer and fewer samples which impacts the power of our tests. Although weighting approaches are common in missing data models \citep{li2013weighting}, an interesting direction for future work is to develop semiparametric methods to use data more efficiently. 

\begin{algorithm}[!t]
	\caption{\textproc{Testing sequential MAR} {\small $(\prec, \mathcal{M}, \mathcal{D}_n)$}}  \label{alg:seq-mar}
	\begin{algorithmic}[1] 
		
		\State Let $\prec$ index variables by $k = 1, \ldots, K.$
		
		\vspace{0.2em}
		\State Let $W_K(\beta^o_K) \coloneqq p(R_K | R_{\prec K}, X^*_{\prec K}; \beta^o_K)$. 
		\State  Estimate $\beta^o_K$ (denote it by $\widehat{\beta}^o_K$).
		
		\vspace{0.25em}
		\For{$k \in \{K-1, \ldots, 1\}$} 
		
		\vspace{0.2em}
		\State Let $W_k(\beta^o_k) \coloneqq p(R_k | R_{\prec k}, X^*_{\prec k}; \beta^o_k)$ and
		
		\hspace{0.4cm} $W_k(\beta^a_k) \coloneqq p(R_k | R_{\prec k}, X^*_{\prec k}, X_{\succ k}; \beta^a_k)$.  
		
		\vspace{0.2em}
		\State Estimate $\beta^o_k$ (denote it by $\widehat{\beta}^o_k$).
		
		\vspace{0.2em}
		\State  Estimate $\beta^a_k$ via the weighted estimating equation:
		{\small
			\begin{align*}
				\mathbb{P}_n \bigg[  \frac{\mathbb{I}(R_{\succ k} = 1)}{\prod_{j \succ k}^{K} W_j(\widehat{\beta}^o_j)}  \times U(\beta^a_k)  \bigg] = 0,
			\end{align*}
		}%
		where $\mathbb{P}_n\big[ U(\beta^a_k) \big] = 0$ is an unbiased estimating equation for $\beta^a_k$ wrt the full law (denote it by $\widehat{\beta}^a_k$). 
		
		\vspace{0.2em}
		\State Compute a weighted likelihood-ratio as follows: 
		{\small
			\begin{align*}
				\rho = n\mathbb{P}_n \bigg[ \frac{\mathbb{I}(R_{\succ k} = 1)}{\prod_{j \succ k}^{K} W_j(\widehat{\beta}^o_j)}   \times \log\Big(  \frac{W_k(\widehat{\beta}^a_k)}{W_k(\widehat{\beta}^o_k)}   \Big)  \bigg]. 
			\end{align*}
		}%
		
		\State Test $\rho$ with $\alpha$ significance level. 
		
		\vspace{0.1cm}
		\If{$\mathcal{M}_o$ is rejected {\small (i.e., $R_k \not\ci X_{\succ k} | R_{\prec k}, X^*_{\prec k}$)}}
		\State \textbf{return} not sequential MAR 
		\EndIf
		
		\EndFor
		\State \textbf{return} sequential MAR
	\end{algorithmic}
\end{algorithm}

\subsection{Sequential MNAR models}

\begin{figure}[!t] 
	\begin{center}
		\scalebox{0.68}{
			\begin{tikzpicture}[>=stealth, node distance=1.4cm]
				\tikzstyle{format} = [thick, circle, minimum size=1.0mm, inner sep=0pt]
				\tikzstyle{square} = [draw, thick, minimum size=4.5mm, inner sep=3pt]
				
				\begin{scope}[xshift=0.cm]
					\path[->, thick]
					node[format] (x11) {$X_1$}
					node[format, right of=x11, xshift=0.45cm] (x21) {$X_2$}
					node[format, right of=x21, xshift=0.45cm] (x31) {$X_3$}
					node[format, below of=x11] (r1) {$R_1$}
					node[format, below of=x21] (r2) {$R_2$}
					node[format, below of=x31] (r3) {$R_3$}
					node[format, below of=r1, yshift=-0.25cm] (x1) {$X^*_1$}
					node[format, below of=r2, yshift=-0.25cm] (x2) {$X^*_2$}
					node[format, below of=r3, yshift=-0.25cm] (x3) {$X^*_3$}
					
					(x11) edge[blue] (x21) 
					(r1) edge[blue] (r2)
					(x1) edge[red, dashed] (r2)
					
					(x31) edge[blue] (r2)
					(x31) edge[blue] (r1)
					(x21) edge[blue] (r1)
					
					(x21) edge[blue] (x31) 
					(r2) edge[blue] (r3)
					(x2) edge[red, dashed] (r3)
					
					(x11) edge[blue, bend left] (x31) 
					(r1) edge[blue, bend right=17] (r3)
					(x1) edge[red, dashed] (r3)
					
					(x11) edge[gray, bend right=25] (x1)
					(x21) edge[gray, bend left=25] (x2)
					(x31) edge[gray, bend left=25] (x3)
					(r1) edge[gray] (x1)
					(r2) edge[gray] (x2)
					(r3) edge[gray] (x3)
					
					node[format, below of=x2, xshift=0cm, yshift=0.75cm] (a) {(a)}; 
				\end{scope}
				
				\begin{scope}[xshift=5.6cm]
					\path[->, thick]
					node[format] (x11) {$X_1$}
					node[format, right of=x11, xshift=0.85cm] (x21) {$X_2$}
					node[format, right of=x21, xshift=0.85cm] (x31) {$X_3=X^*_3$}
					node[format, below of=x11] (r1) {$R_1$}
					node[format, below of=x21] (r2) {$R_2$}
					node[square, below of=x31] (r3) {$R_3=1$}
					node[format, below of=r1, yshift=-0.25cm] (x1) {$X^*_1$}
					node[format, below of=r2, yshift=-0.25cm] (x2) {$X^*_2$}
					node[format, below of=r3, yshift=-0.25cm] (x3) {}
					
					(x11) edge[blue] (x21) 
					(r1) edge[blue] (r2)
					(x1) edge[red, dashed] (r2)
					
					(x31) edge[blue] (r2)
					(x31) edge[blue] (r1)
					(x21) edge[blue] (r1)
					
					(x21) edge[blue] (x31) 
					
					(x11) edge[blue, bend left] (x31) 
					
					(x11) edge[gray, bend right=25] (x1)
					(x21) edge[gray, bend left=25] (x2)
					(r1) edge[gray] (x1)
					(r2) edge[gray] (x2)
					
					node[format, below of=x2, xshift=0cm, yshift=0.75cm] (b) {(b)}; 
				\end{scope}
				
			\end{tikzpicture} 
		}
		\vspace{-0.2cm}
		\caption{ (a) Example of a sequential MNAR model (without the dashed edge) along with its permutation supermodel (with the dashed edge); (b) The graph Markov wrt the intervention distribution $p(. | \text{do}(R_3=1))$.} 
		\label{fig:seq-mnar}
		\vspace{-0.6cm}
	\end{center}
\end{figure}
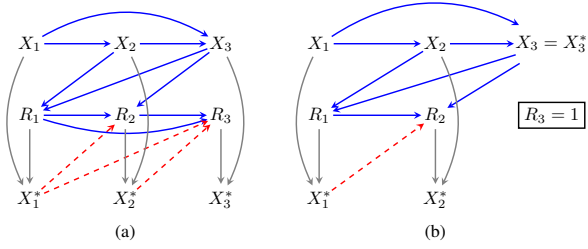

We call a missing data model a \textit{sequential MNAR} model if under an ordering $\prec$ that indexes variables by $k = 1, \ldots, K$, the following set of independence restrictions hold: 
\begin{align}
	\hspace{-0.325cm}	R_k \ci X_{\prec k+1}, X^*_{\prec k} \mid R_{\prec k}, X_{\succ k}, \forall k \hspace{0.15cm} \textit{\small (sequential-MNAR)}  \hspace{-0.15cm}
	\label{eq:no-colluder}
\end{align}%
An example of this model is shown in Fig.~\ref{fig:seq-mnar}(a) (without the dashed edges.) We can view the sequential MNAR model as a submodel of the permutation model since in addition to the restrictions in (\ref{eq:perm}), it assumes $R_k \ci X^*_{\prec k} \mid R_{\prec k}, X_{\succ k}, \forall k$.  Thus, we focus on testable implications of these extra assumptions and propose ways to empirically evaluate them. 

Unlike sequential MAR models, the d-separation statements being tested in sequential MNAR models are between missingness indicators and proxy variables, which can be viewed as context-specific restrictions. Due to determinism, when $R_j = 0, X^*_j=``?"$, an independence restriction such as $R_k \ci X^*_j | R_j=0$ becomes a statement of independence between a random variable $R_k$ and some constant, which is trivially true. Hence, the set  $R_k \ci X^*_{\prec k} | R_{\prec k}, X_{\succ k}, \forall k$  is equivalent to  context-specific restrictions $R_k \ci X_{\prec k} | R_{\prec k}=1,  X_{\succ k}, \forall k$; note that  $R_{\prec k}$ is evaluated at one. 
Even though, these independences
restrict us to rows where $X_{\prec k}$ is fully observed, we still need enough assumptions to  plug in $R_{\succ k} = 1$ in the conditioning set, since $X_{\succ k}$ is in the conditioning set. Unfortunately, the independence between $R_k$ and $X^*_{\prec k}$ no longer holds if we condition on $R_{\succ k}$. However, the following theorem formalizes that restrictions in sequential MNAR models defined above can still be tested as Verma constraints in identified intervention distributions where $R_{\succ k}$ are intervened and $X_{\succ k}$ are fully observed. 
\begin{theorem} \label{thm:seq-mnar} 
	The independence $R_k \ci X^*_{\prec k} | R_{\prec k}, X_{\succ k}$ has a testable implication on the observed data distribution in the form of a  Verma constraint $R_k \ci X^*_{\prec k} | R_{\prec k}, X_{\succ k}, \text{do}(R_{\succ k}  = 1)$, where the intervention distribution $p(X, R \setminus R_{\succ k}, X^* | \text{do}(R_{\succ k} =1))$ is identified. 
\end{theorem}
To evaluate these Verma constraints, we use weighted likelihood-ratio tests again. We explain this via a sequential MNAR example in Fig.~\ref{fig:seq-mnar}(a) (without the dashed edges.) ${\cal M}_o$ is the statistical model of this DAG and ${\cal M}_a$ is the permutation supermodel with the dashed edges. We are interested in testing absence of the dashed edges which imply $R_3 \ci X^*_1, X^*_2 \mid R_1, R_2$ and $R_2 \ci X^*_1 \mid R_1, X_3.$ To empirically evaluate the first restriction, we need to compare $p(R_3 \mid R_1, R_2, X^*_1, X^*_2)$ and $p(R_3 \mid R_1, R_2),$ which is straightforward  since  these two models are  direct functions of observed data. To evaluate the second restriction however, we need to compare $p(R_2 | X^*_1, R_1, X_3)$ and $p(R_2 | R_1, X_3)$ wrt the intervention distribution $p(. | \text{do}(R_3=1)$, which  corresponds to the truncated factorization $p(X, R, X^*) / p(R_3| R_1, R_2, X^*_1, X^*_3)$ (evaluated at $R_3=1$) and is Markov relative to the graph in Fig.~\ref{fig:seq-mnar}(b).  Thus, we can use $p(R_3 | R_1, R_2, X^*_1, X^*_2)$ as inverse weights to fit models wrt this truncated distribution. Let $W_{r_2}(\beta^a_{r_2}) \coloneqq p(R_2 | R_1, X^*_1, X_3; \beta^a_{r_2})$ and let $\mathbb{P}_n[U(\beta^a_{r_2})] = 0$ be an unbiased estimating equation for $\beta^a_{r_2}$ wrt the full law. We can estimate $\beta^a_{r_2}$ using observed data via this weighted estimating equation:  $\mathbb{P}_n[  \{R_3 / p(R_3 | R_1, R_2, X^*_1, X^*_2; \widehat{\eta})\} \times U(\beta^a_{r_2}) ] = 0,$ where $\widehat{\eta}$ is the estimated parameters for $p(R_3 | R_1, R_2, X^*_1, X^*_2)$. Following the same logic, we can also estimate ${\beta}^o_{r_2}$ in $W_{r_2}(\widehat{\beta}^o_{r_2}) \coloneqq p(R_2 | R_1, X_3; \beta^o_{r_2} )$. Finally, we use the following statistic in a weighted likelihood-ratio to test the restriction $R_2 \ci X^*_1 \mid R_1, X_3$: 

\vspace{-0.6cm}
{\small
	\begin{align*}
		\rho = n\mathbb{P}_n\bigg[ \frac{R_3}{p(R_3 \mid R_1, R_2, X^*_1, X^*_2; \widehat{\eta})} \times \log\Big( \frac{ W_{r_2}(\widehat{\beta}^a_{r_2})}{W_{r_2}(\widehat{\beta}^o_{r_2})} \Big) \bigg]. 
	\end{align*}
}%
If we test the restriction $R_3 \ci X^*_1, X^*_2 | R_1, R_2$ first and conclude that the independence  holds, we can use the $R_3$ fitted propensity score under the accepted null, that is $p(R_3 \mid R_1, R_2; \widehat{\beta}^o_{r_3})$, in above (without conditioning on $X^*_1, X^*_2$). This implies that for testing $R_2 \ci X^*_1 \mid R_1, X_3$, we do not have to use  the full permutation model as a supermodel. Instead, we can use the permutation model where the $\{X^*_1, X^*_2\} \rightarrow R_3$ edges are absent. 

Algorithm~1 in Appendix~B.1 provides an automated procedure for performing sequential goodness-of-fit tests based on weighted likelihood-ratios for $K \! > \! 3$ variables. The algorithm is similar to testing sequential MAR, but due to space limits, it is defferred to the supplements. 

{\bf Remark 1. } 
It is worth pointing out that the sequential MNAR model is a special case of  models Markov relative to m-DAGs with no \textit{colluders} studied in \cite{nabi20completeness} -- a colluder exists at $R_j$ if there exists $X_i \in X \setminus X_j$ such that $X_i \rightarrow R_j \leftarrow R_i$.  \cite{nabi20completeness} 
showed that under the absence of colluder structures and self-censoring edges ($X_k \rightarrow R_k$), the full law Markov relative to such an m-DAG is identified. Further, they showed that such m-DAGs are a submodel of the saturated no self-censoring model defined in (\ref{eq:no-self}). Thus, it is possible to use the no self-censoring model as an alternative supermodel to test some of the restrictions in the sequential MNAR model. Namely, we can empirically evaluate this set of restrictions: $R_k \ci X_{\prec k} \mid R_{-k}, X_{\succ k}.$ 

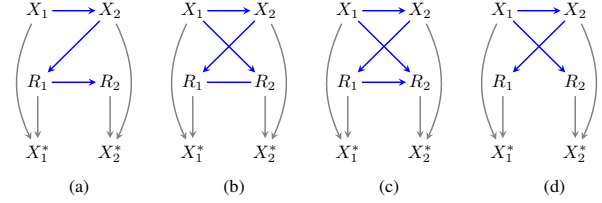
\begin{figure}[!t] 
	\begin{center}
		\scalebox{0.68}{
			\begin{tikzpicture}[>=stealth, node distance=1.4cm]
				\tikzstyle{format} = [thick, circle, minimum size=1.0mm, inner sep=0pt]
				\tikzstyle{square} = [draw, thick, minimum size=4.5mm, inner sep=3pt]
				
				\begin{scope}[xshift=0cm, yshift=0cm]
					\path[->, thick]
					node[format] (x11) {$X_1$}
					node[format, right of=x11, xshift=-0.0cm] (x21) {$X_2$}
					node[format, below of=x11] (r1) {$R_1$}
					node[format, below of=x21] (r2) {$R_2$}
					node[format, below of=r1] (x1) {$X^*_1$}
					node[format, below of=r2] (x2) {$X^*_2$}
					
					(x11) edge[blue] (x21) 
					(r1) edge[blue] (r2)
					(x21) edge[blue] (r1)
					
					(x11) edge[gray, bend right=25] (x1)
					(x21) edge[gray, bend left=25] (x2)
					(r1) edge[gray] (x1)
					(r2) edge[gray] (x2)
					
					node[format, below of=x1, xshift=0.8cm, yshift=0.75cm] (a) {(a)} ;
				\end{scope}	
				
				\begin{scope}[xshift=3.cm, yshift=0cm]
					\path[->, thick]
					node[format] (x11) {$X_1$}
					node[format, right of=x11, xshift=-0.0cm] (x21) {$X_2$} 
					node[format, below of=x11] (r1) {$R_1$}
					node[format, below of=x21] (r2) {$R_2$}
					node[format, below of=r1] (x1) {$X^*_1$}
					node[format, below of=r2] (x2) {$X^*_2$}
					
					(x11) edge[blue] (x21) 
					(x11) edge[blue] (r2)
					(x21) edge[blue] (r1)
					(r1) edge[blue, -] (r2)
					
					(x11) edge[gray, bend right=25] (x1)
					(x21) edge[gray, bend left=25] (x2)
					(r1) edge[gray] (x1)
					(r2) edge[gray] (x2)
					
					node[format, below of=x1, xshift=0.8cm, yshift=0.75cm] (b) {(b)} ;
				\end{scope}
				
				\begin{scope}[xshift=6cm, yshift=0cm]
					\path[->, thick]
					node[format] (x11) {$X_1$}
					node[format, right of=x11, xshift=-0.0cm] (x21) {$X_2$}
					node[format, below of=x11] (r1) {$R_1$}
					node[format, below of=x21] (r2) {$R_2$}
					node[format, below of=r1] (x1) {$X^*_1$}
					node[format, below of=r2] (x2) {$X^*_2$}
					
					(x11) edge[blue] (x21) 
					(x11) edge[blue] (r2)
					(x21) edge[blue] (r1)
					(r1) edge[blue] (r2)
					
					(x11) edge[gray, bend right=25] (x1)
					(x21) edge[gray, bend left=25] (x2)
					(r1) edge[gray] (x1)
					(r2) edge[gray] (x2)
					
					node[format, below of=x1, xshift=0.8cm, yshift=0.75cm] (c) {(c)} ;
				\end{scope}
				
				\begin{scope}[xshift=9cm]
					\path[->, thick]
					node[format] (x11) {$X_1$}
					node[format, right of=x11, xshift=0.cm] (x21) {$X_2$}
					node[format, below of=x11] (r1) {$R_1$}
					node[format, below of=x21] (r2) {$R_2$}
					node[format, below of=r1] (x1) {$X^*_1$}
					node[format, below of=r2] (x2) {$X^*_2$}
					
					(x11) edge[blue] (x21) 
					(x11) edge[blue] (r2)
					(x21) edge[blue] (r1)
					
					(x11) edge[gray, bend right=25] (x1)
					(x21) edge[gray, bend left=25] (x2)
					(r1) edge[gray] (x1)
					(r2) edge[gray] (x2)
					
					node[format, below of=x1, xshift=1cm, yshift=0.75cm] (d) {(d)} ;
				\end{scope}
				
			\end{tikzpicture} 
		}
		\vspace{-0.2cm}
		\caption{The sequential MNAR  model in (a) can be tested as a submodel of the saturated no self-censoring model in (b); (c) A criss-cross supermodel of (a) where the test statistic is not identifiable; (d) Example of a block-parallel MNAR model which can be tested as a submodel of (b). } 
		\label{fig:seq-mnar_2}
		\vspace{-0.6cm}
	\end{center}
\end{figure}

As an example, consider the absence of an edge between $X_1$ and $R_2$ in Fig.~\ref{fig:seq-mnar_2}(a) which implies $R_2 \ci X_1 | R_1$. 
The no self-censoring supermodel is drawn in Fig.~\ref{fig:seq-mnar_2}(b) (with $R_1, R_2$ edge undirected). We can evaluate this independence by showing $p(R_2 | R_1, X_1)$ is not a function of $X_1$. See Appendix~B.2 for  details on how to set up such a test. 


{\bf Remark 2.}  
The m-DAG in Fig.~\ref{fig:seq-mnar_2}(c) is also a supermodel of Fig.~\ref{fig:seq-mnar_2}(a). However, we cannot use it to evaluate the independence $R_2 \ci X_1 \mid R_1$, because $p(R_2 | R_1, X_1)$ 
is not fully identified under this supermodel, due to the colluder structure at $R_2$ as shown by \cite{bhattacharya19mid}. 

We call the m-DAG in Fig.~\ref{fig:seq-mnar_2}(c), the \textit{criss-cross} structure. In the following theorem, we show that unlike the permutation and no self-censoring models, the target law (and thus the full law) is not identified when such structures are present. 
\begin{theorem} \label{thm:criss-cross} 
	The target law $p(X)$ is not identified in an m-DAG model where there exists at least one criss-cross structure between a pair of variables. 
\end{theorem}
The above result characterizes a novel graphical structure 
that impedes target law identification; this may lead to further insights on an open problem regarding the discovery of a \textit{sound} and \textit{complete} algorithm for target law identification. 

\textbf{Remark 3.} 
As an alternative to the likelihood-ratio test, we can compute odds ratios to perform independence tests. For instance, in the MAR model of Fig.~\ref{fig:seq-mar}(a), $R_1 \ci X_2$ translates into $\text{OR}(R_1, X_2) = 1$, and in the MNAR model of Fig.~\ref{fig:seq-mnar_2}(a), $R_2 \ci X^*_1 | R_1$ translates into $\text{OR}(R_2, X_1 | R_1 = 1) = 1$, which both can be empirically evaluated. See Appendices~C.2 and C.3 for a generalization of the idea of using odds ratio for goodness-of-fit tests in the sequential MAR and MNAR models with $K > 2$ variables. 



\subsection{Block parallel MNAR models}

We call a missing data model a \textit{block-parallel MNAR} model\footnote{Block-parallel model was introduced in \cite{mohan2013missing}.} if it satisfies the following set of independence restrictions: 
\begin{align}
	R_k \ci R_{-k}, X_k \mid X_{-k}, \forall k  \ \textit{ (block-parallel MNAR)}
	\label{eq:block-par}
\end{align}%
An example of this model is shown in Fig.~\ref{fig:seq-mnar_2}(d). Using graphoid axioms, it is easy to show that the block-parallel model assumes $R_k \ci R_j \mid X, \forall j\not=k$ on top of what the no self-censoring model, defined in (\ref{eq:no-self}),  already assumes. Thus, we view the block-parallel model (${\cal M}_o$) as a submodel of the saturated no self-censoring model (${\cal M}_a$), and focus on testable implications and  empirical evaluations of these extra assumptions. Unlike the sequential models, the independence statements here are between missingness indicators and there is no predefined ordering. 

If we were to follow ideas from the previous two subsections, we would need to intervene on $R_k$ and $R_j$ to test the independence $R_k \ci R_j \mid X$ as $X_k, X_j$ appear in the conditioning set. Interventions on $R_k$ and $R_j$ fix them to constants, which prevent us from evaluating  independence. One might then conclude that such constraints are untestable. However, we use odds-ratio parameterization of the missingness mechanism to argue that these restrictions are indeed testable. We formalize the  results in the following theorem. 
\begin{theorem}
	The independence $R_k \ci R_j |  X$ $\forall j \not=k$ has a testable implication on observed data law which can be stated via $\text{OR}(R_k, R_j | X_{-kj}, R_{-kj} = 1) = 1$.
	\label{thm:block-par}
\end{theorem}
%
As an example, consider 
the m-DAG in 
Fig.~\ref{fig:seq-mnar_2}(d) 
and its 
 supermodel in (b).  The absence of an edge between $R_1, R_2$ implies $R_1 \! \ci \! R_2 | X.$ This is equivalent to stating that the odds ratio between $R_1$ and $R_2$ conditioned on $X_1, X_2$ is one, i.e.,  $\text{OR}(R_1=0, R_2=0 | X) = 1$. See Appendix~A.1 for a description of the odds ratio parameterization. 
Let $\theta$ denote the odds ratio. \cite{malinsky2021semiparametric} proposed the following unbiased estimating equation to estimate $\theta$:

\vspace{-0.5cm}
{\small
	\begin{align*}
		\mathbb{P}_n \Big[ R_1 R_2 \times \frac{p(R_1=0, R_2=0 \mid X)}{p(R_1 = 1, R_2=1 \mid X)} \! - (1 \! - \! R_1) (1 \! - \! R_2) \Big] \!\!\!= 0, 
	\end{align*}
}%
where $\theta$ appears in the density ratio since it equals:
{\small
	\begin{align*}
		\frac{p(R_1  =0 \mid R_2 = 1, X_2) }{p(R_1  =1 \mid R_2 = 1, X_2) }  \times \frac{p(R_2 = 0 \mid R_1 = 1, X_1)}{p(R_2 = 1 \mid R_1 = 1, X_1)} \times \theta.
	\end{align*}
}%
See Appendix~A.1.1 for detailed derivations. For $K \! > \! 2$ variables, we can test the absence of edges between any two pairs of missingness indicators by computing pairwise odds ratios. We formalize the goodness-of-fit tests based on these calculations in Algorithm~2 outlined in  Appendix~C.4. 

\subsection{Extensions to settings with unmeasured confounders}
\label{sec:unmeasured_confounding}

We can extend the applicability of our results to scenarios where not only variables are missing but some are completely unobserved, by considering hidden variable DAGs ${\cal G}(V \cup U)$, where $V = \{X, R, X^*\}$ and the variables in $U$ are unobserved. 
In such cases, we can obtain a missing data acyclic directed mixed graph (m-ADMG) ${\cal G}(V)$, by applying the latent projection operator \citep{verma1990equivalence} to the hidden variable DAG ${\cal G}(V \cup U)$. The full law then follows the nested Markov factorization \citep{richardson2023nested} with respect to the m-ADMG ${\cal G}(V)$. 
%
A m-ADMG obtained through the projection of a hidden variable m-DAG adheres to the same edge restrictions.

If there is a concern about latent confounding in the target law, our framework allows for arbitrary confounding among the $X$ variables without any modifications to the proposed tests in the previous section. This means we can incorporate unmeasured confounders of the form $X_i \leftarrow U \rightarrow X_j$, often represented as a bidirected edge $X_i \leftrightarrow X_j$ between any pair of variables $X_i$ and $X_j$. It is important to note that the missing data mechanism $p(R|X)$ is largely unaffected by the inclusion of these bidirected edges. For example, in Fig.~\ref{fig:seq-mar2}(a), we can introduce unmeasured confounders between every pair of variables in $X$ while maintaining the same testable implications and goodness-of-fit tests. 

\begin{figure*}[!t]
	\centering
	\begin{subfigure}{.5\textwidth}
		\centering
		\includegraphics[scale=0.32]{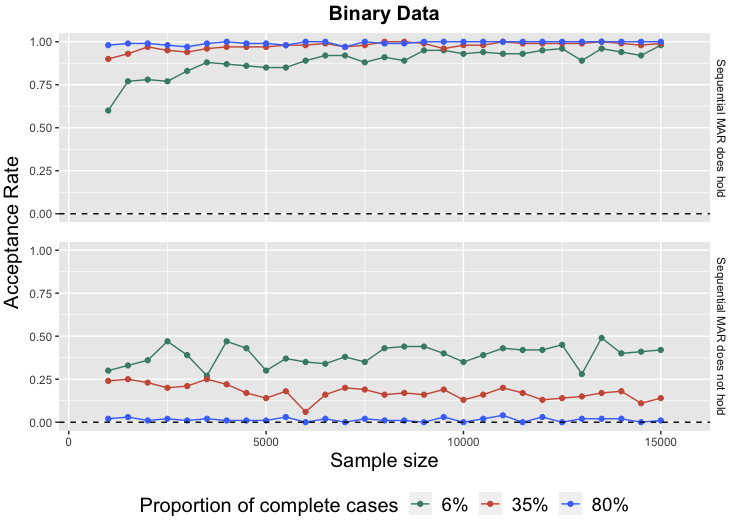}
		\label{fig:mar_bin}
	\end{subfigure}%
	\begin{subfigure}{.5\textwidth}
		\centering
		\includegraphics[scale=0.32]{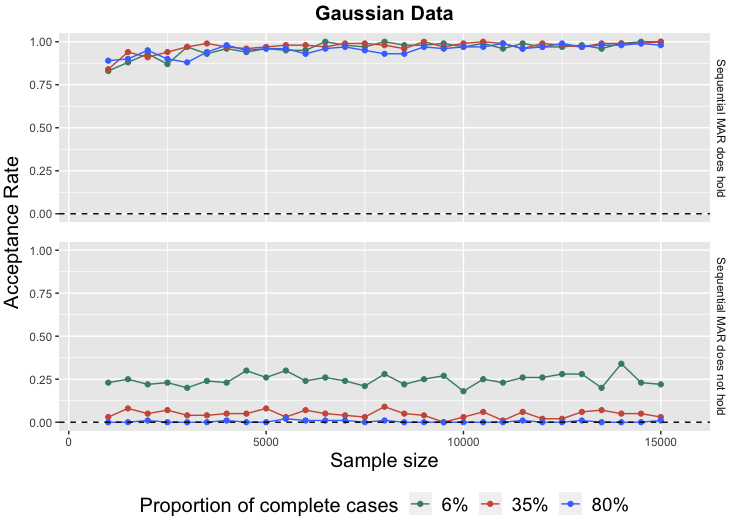}
		\label{fig:mar_cont}
	\end{subfigure}
	\caption{Results on testing \textbf{sequential MAR} models. \textit{(top row)} The sequential MAR model captures the true underlying missingness mechanism. \textit{(bottom row)} The assumptions of sequential MAR model are violated. } 
	\label{fig:mar_sim}
\end{figure*}

Unmeasured confounding is also possible in the missing data mechanism, as long as there are no ``colluding paths'' between any $X_i$ and its corresponding missingness indicator $R_i$. A colluding path is a path where every node on the path is a collider. The presence of colluding paths leads to the non-identification of the full law, potentially affecting the validity of the proposed goodness-of-fit tests. For instance, in Fig.~\ref{fig:seq-mnar}(a), we can have unmeasured confounders between $X_3$ and $R_2$, or between $X_2$ and $R_1$, or between all variables in $X$, and so on. For a formal definition of colluding paths and a detailed explanation of why they result in non-identification, refer to \citep{nabi20completeness}.

\section{Simulations}
\label{sec:sims} 

We conduct three sets of simulation analyses to illustrate the utility of our proposed methods in testing the missing data restrictions using only partially observed samples. Each set focuses on a class of m-DAGs that were considered here. For each simulation, we generate four random variables from either a multivariate normal distribution or  binomial distribution. We induce missing values in all four variables according to a missingness mechanism that follows restrictions of either sequential MAR, sequential MNAR, block-parallel, or supermodels of them. The exact data generating mechanism is described in Appendix~E.  R code can be found at \url{https://github.com/raziehna/missing-data-testability}. 




In the main body, 
we 
present results 
on testing the sequential MAR model defined via the set of restrictions in (\ref{eq:seq-mar}). We follow Algorithm~\ref{alg:seq-mar} to test the independence restrictions, which entails running a total of $K-1$ tests. Our test statistic is $2\rho$ and we use a chi-square distribution with $K-k$ degrees of freedom to evaluate the goodness-of-fits --  the degree of freedom is chosen as the difference between number of parameters in $W_k(\beta^a_k)$ and $W_k(\beta^0_k)$, as defined in the algorithm. If the p-values are all greater than $0.05$, we accept the sequential MAR model. 
Results on sequential MNAR and block-parallel MNAR models are provided in Appendix~E.  

For a fixed sample size, we simulate $100$ different datasets and calculate the acceptance rate of a sequential MAR model. The acceptance rate is plotted as a function of sample size in Fig.~\ref{fig:mar_sim}. The sample size ranges from $1,000$ to $15,000$ with $500$  increments. In each panel, there are three plots that vary in terms of the proportion of complete cases in the dataset, i.e., $6\%, 35\%, 80\%$ which is achieved by changing the range in the uniform distribution where the parameters are sampled from (the proportion of complete cases is taken as an average of complete cases over $100$ iterations). The top row of Fig.~\ref{fig:mar_sim} illustrates the results when the true underlying missingness mechanism satisfies the assumptions of the sequential MAR model, 
and the bottom row illustrates results for when the restrictions are no longer valid 
As seen in the figure, the acceptance rate is quite high when the 
sequential MAR 
model holds true and it is low when the 
model does not hold, even if we have only  $6\%$ complete cases which is impressive performance with small data. The plots at the bottom row also illustrate that  the tests would perform better in terms of rejecting the sequential MAR model while the truth is not MAR when the missingness rate decreases; with $80\%$ complete cases the acceptance rate vanishes.

\section{Conclusions}
\label{sec:conc} 

Independence restrictions in a missing data model might be empirically untestable, or they might translate into more complex restrictions on the observed data law than ordinary d-separation statements. In this paper, we considered various graphical models of missing data and investigated testable implications of the  underlying statistical assumptions on the observed data law. We have extended the notion of testability in prior literature by viewing ordinary conditional independence tests as Verma constraints in intervention distributions. 
We have proposed goodness-of-fit tests based on weighted likelihood-ratio tests and 
odds-ratio parameterizations. 
Our results 
are essential 
in validating the assumed statistical missing data models 
in practice and discovering the mechanisms that drive the missingness of variables. A potential future direction is to develop estimation methods that would complement our proposals by allowing a more efficient use of data in 
performing goodness-of-fit tests. 


%

\begin{acknowledgements} 
	This work is partly supported by the National Center for Advancing Translational Sciences of the National Institutes of Health under Award Number UL1TR002378. 
\end{acknowledgements}

\clearpage 

\bibliography{references}

\clearpage 

\appendix

\onecolumn 

{\Large \bf APPENDIX}

\vspace{0.25cm}

In Appendix~\ref{app:basics}, we cover additional preliminaries: (i) we present the odds-ratio parameterization of a missing data process and demonstrate the estimation of an odds ratio through a straightforward example, (ii) we elaborate more on parameter counting in discrete models to assess whether the assumptions in a full law impose restrictions on observed data law, and (iii) we provide additional details on substantive edge distinctions between $\{V^*_i, R_j\}$ vs $\{V_i, R_j\}$ in the permutation model.  
Appendix~\ref{app:likelihood-tests} contains additional discussions on the goodness-of-fit tests in the sequential MNAR model using likelihood approaches. It also includes an automated algorithm for performing a sequential goodness-of-fit tests based on weighted likelihood-ratios. 
Appendix~\ref{app:odds_test} contains additional discussions on the use of odds-ratio parameterization in the sequential MAR and sequential MNAR models, as well as a formalization of the goodness-of-fit tests in block-parallel MNAR models based on odds ratio calculations. Appendix~\ref{app:proof} contains the proofs. Appendix~\ref{app:sims} contains simulation details and additional empirical analyses. 

\appendix 

\section{Preliminaries}
\label{app:basics}

\subsection{Odds-ratio parameterization}
\label{app:odds}

The odds-ratio parameterization of joint distributions  $p(R | X)$ was introduced in \cite{chen2007semiparametric}.  Assuming we have $K$ missingness indicators, $p(R \mid X)$ can be expressed as follows: 
{\small
	\begin{align}
		p(R \mid X) 
		= \ \frac{1}{Z}\times \prod_{k = 1}^{K} \ p(R_k \mid R_{-k} = 1,X)  
		\times \prod_{k = 2}^{K} \text{OR}(R_k, R_{\prec k} \mid R_{\succ k} = 1, X), 
		\label{eq:odds_ratio_chen}
	\end{align}
}%
where $R_{-k} = R \setminus R_k, R_{\prec k} = \{R_1, \ldots, R_{k - 1}\}, R_{\succ k} = \{R_{k+1}, \ldots, R_K\}$, and 
{\small
	\begin{align*}
		&\text{OR}(R_k, R_{\prec k} \mid R_{\succ k} = 1, X) 
		= \frac{p(R_k \mid R_{\succ k} = 1, R_{\prec k}, X)}{p(R_k = 1 \mid R_{\succ k} = 1, R_{\prec k}, X)} 
		\times 
		\frac{p(R_k = 1 \mid R_{-k} =1, X)}{p(R_k \mid R_{-k} = 1, X)}.
	\end{align*}
}%
$Z$ in Eq.~(\ref{eq:odds_ratio_chen}) is the normalizing term and is equal to {\small$ \sum_{r} \Big\{ \prod_{k = 1}^{K} \ p(r_k \mid R_{-k} = 1,X) \times \prod_{k = 2}^{K} \text{OR}(r_k, r_{\prec k} \mid R_{\succ k} = 1, X) \Big\}$}.

\vspace{0.25cm}
{\bf Estimating equations for computing odds ratios. } 

Consider the no self-censoring model with two variables, shown in Fig.~\ref{fig:seq-mnar_2}(b). Let $\theta(r_1, r_2) =  \text{OR}(R_1 = r_1, R_2=r_2 \mid X_1, X_2)$. We can estimate $\theta(r_1=0, r_2=0)$ with the following unbiased estimating equation where an odds-ratio parameterization of $p(R | X)$ is used in place. We have: 
\begin{align*}
	p(R_1=r_1, R_2=r_2 \mid X) = \frac{1}{Z} \times p(R_1 = r_1 | R_2 =1, X_2) \times p(R_2=r_2 | R_1=1, X_1) \times \theta(r_1, r_2). 
\end{align*}
Therefore,  
{\small
	\begin{align*}
		&\mathbb{P}_n \Big[ R_1 R_2 \times \frac{p(R_1=0, R_2=0 \mid X)}{p(R_1 = 1, R_2=1 \mid X)} -  (1 - R_1) (1  -  R_2) \Big] \\
		&\hspace{1cm}  = \mathbb{P}_n \Big[ R_1 R_2 \times \frac{p(R_1 = 0| R_2 =1, X_2) \times p(R_2=0 | R_1=1, X_1) \times \theta(R_1=0, R_2=0)}{p(R_1 = 1| R_2 =1, X_2) \times p(R_2=1 | R_1=1, X_1) \times \theta(R_1=1, R_2=1)} -  (1 - R_1) (1  -  R_2) \Big]  \\
		&\hspace{1cm}  = \mathbb{P}_n \Big[ R_1 R_2 \times \frac{p(R_1 = 0| R_2 =1, X_2) \times p(R_2=0 | R_1=1, X_1) }{p(R_1 = 1| R_2 =1, X_2) \times p(R_2=1 | R_1=1, X_1) } \times  \theta(R_1=0, R_2=0) -  (1 - R_1) (1  -  R_2) \Big] \\
		&\hspace{1cm}  = 0. 
	\end{align*}
}%
The first equality holds by definition, the second equality holds because $\text{OR}(R_1=1, R_2=1) = 1$, and the third equality can be simply proved with tower laws of expectations. 
Given the above, we can find a closed form estimator for $ \theta(R_1=0, R_2=0)$: 
{\small
	\begin{align*}
		\theta(R_1=0, R_2=0) = \frac{\mathbb{P}_n \Big[ (1-R_1) \times (1-R_2) \Big]}{ \mathbb{P}_n \bigg[ R_1 \times R_2 \times \displaystyle \frac{p(R_1 = 0| R_2 =1, X_2) \times p(R_2=0 | R_1=1, X_1) }{p(R_1 = 1| R_2 =1, X_2) \times p(R_2=1 | R_1=1, X_1) } \bigg] }. 
	\end{align*}
}

For $K > 2$, we need to compute odds ratio terms of the form $\theta(R_k=0, R_j=0)  \coloneqq \text{OR}(R_k = 0, R_j = 0 | R_{-kj}=1, X)$. The following unbiased estimating equation that incorporates $R_{-kj}$ can be used to estimate $\theta(R_k=0, R_j=0)$: 
{\small
	\begin{align*}
		&\mathbb{P}_n \bigg[ \prod_{i=1}^K R_i \times \frac{p(R_k = 0| R_{-k} =1, X_{-k}) \times p(R_j=0 | R_{-j}=1, X_{-j})}{p(R_k = 1| R_{-k} =1, X_{-k}) \times p(R_j=1 | R_{-j}=1, X_{-j})} \times  \theta(R_k=0, R_j=0)  - \prod_{i \not= \{j, k\}} R_i (1 - R_k) (1  -  R_j) \bigg]  = 0. 
	\end{align*}
}%
Using the tower laws of expectations, it is easy to show why the above estimating equation holds.

\subsection{Parameter counting argument} 
\label{app:par_count}

How does one know that a missing data DAG imposes restrictions that are testable from the observed data distribution? When all substantive variables take on values in a finite discrete state space, one simple check is to compare the number of parameters in the full law using the DAG factorization in (\ref{eq:factor}) and the saturated observed data law using the \emph{pattern-mixture} factorization \citep{rubin76inference}. The pattern-mixture factorization is given by the marginal distribution of $R$ and the conditional distribution of $X^*$ given $R.$ If a missing data DAG with an identified full law can be described with fewer parameters than the saturated pattern-mixture model, we may conclude that the restrictions on full law impose constraints on the observed data distribution. \cite{shpitser2016consistent} has used parameter counting to give an intuition for why the no self-censoring model is identified. \cite{nabi20completeness} also have relied on a parameter counting argument to prove the completeness of their results for full law identification in missing data DAG models. 

As an example, consider a missing data model with two substantive binary variables $X_1$ and $X_2$. Assume the full law satisfies the assumptions of the  permutation model in (\ref{eq:perm}), which are $R_1 \ci X_1 | X_2$ and $R_2 \ci X_1, X_2 \mid R_1, X^*_1$. The full law then factorizes as $p(X_1, X_2) \times p(R_1 | X_2) \times p(R_2 | R_1, X^*_1)$. We need $3$ parameters for parameterizing $p(X_1, X_2),$ $2$ parameters for $p(R_1 \mid X_2),$ and $3$ parameters for $p(R_2 | R_1, X^*_1);$ thus a total of $8$ parameters. (We excluded the deterministic terms $p(X^*_1 | R_1, X_1)$ and $p(X^*_2 | R_2, X_2)$ as they do not add any parameters.) On the other hand, the pattern-mixture factorization of the observed data law $p(R, X^*)$ can be written as $p(R_1, R_2) \times p(X^*_1, X^*_2 | R_1, R_2).$ Since $R_1$ and $R_2$ are binary, it requires at most $3$ parameters to parameterize $p(R_1, R_2).$ Using  chain rule factorization, we have $p(X^* | R) = p(X^*_1 | R_1, R_2) \times p(X^*_2 | R_1, R_2, X^*_1).$ Due to the deterministic relations, if $R_1 = 0$ then $X^*_1 = ``?"$, thus we need at most $2$ parameters to parameterize $p(X^*_1 | R_1, R_2)$. Similarly, we need at most $3$ parameters to parameterize $p(X^*_2 | R_1, R_2, X^*_1).$ In total, $8$ parameters are required to encode a saturated observed data law. As expected, the number of parameters in the full law of the permutation model (which is proven to be identified as a function of observed data) and the saturated observed data law are the same, reaffirming the fact that permutation model is saturated and places no restrictions on the observed data distribution. 

As another example of a saturated model, consider the no self-censoring model in Fig.~\ref{fig:seq-mnar_2}(b). The odds-ratio parameterization of the  missingness mechanism $p(R | X)$ is as  follows: 
\begin{align} 
	&p(R_1 = r_1, R_2 = r_2 \mid X_1, X_2) \label{eq:odds} \\
	&\hspace{1.5cm}= \frac{1}{Z}  \times p(R_1 = r_1 \mid R_2 = 1, X_1, X_2) \times p(R_2 = r_2 \mid R_1 = 1, X_1, X_2) \times \text{OR}(R_1  =r_1, R_2 = r_2 \mid  X_1, X_2)  \nonumber \\
	&\hspace{1.5cm}=  \frac{1}{Z}  \times p(R_1 = r_1 \mid R_2 = 1, X_2) \times p(R_2 = r_2 \mid R_1 = 1, X_1) \times f(R_1  =r_1, R_2 = r_2),  \nonumber 
\end{align}%
where $Z = \sum_{r_1, r_2} p(R_1 = r_1 \mid R_2 = 1, X_2) \times p(R_2 = r_2 \mid R_1 = 1, X_1,) \times \text{OR}(R_1  =r_1, R_2 = r_2 | X_1, X_2)$.  
The second equality in (\ref{eq:odds}) holds because $R_1 \ci X_1 | R_2, X_2$ and $R_2 \ci X_2 | R_1, X_1$. Further, $\text{OR}(R_1  =r_1, R_2 = r_2 \mid  X_1, X_2)$ is just a function of $R_1$ and $R_2$ because: 
\begin{align*}
	\text{OR}(R_1 = r_1, R_2=r_2 \mid X_1, X_2)
	& = \frac{p(R_1=r_1 \mid R_2=r_2, X_2)}{p(R_1 = 1 \mid R_2=r_2, X_2)} \times \frac{p(R_1  = 1 \mid R_2 = 1, X_2)}{p(R_1=r_1 \mid R_2 = 1, X_2)} \\ 
	& = \frac{p(R_2=r_2 \mid R_1=r_1, X_1)}{p(R_2 = 1 \mid R_1=r_1, X_1)} \times \frac{p(R_2  = 1 \mid R_1 = 1, X_1)}{p(R_2=r_2 \mid R_1 = 1, X_1)} \\ 
	&= f(R_1, R_2).
\end{align*}
The first equality holds because $R_1 \ci X_1 \mid R_2, X_2,$ the second equality holds because $R_2 \ci  X_2 \mid R_1, X_1$, and together they imply the last equality which means $\text{OR}(R_1, R_2 \mid X_1, X_2)$ is a function of $R_1, R_2$ (all observed data). In the above argument, we have used the fact that odds ratios is symmetric (i.e., $\text{OR}(A, B | Z) = \text{OR}(B, A | Z)$). Assuming $X_1$ and $X_2$ are binary, the full law in a no self-censoring model would have $8$ parameters (same number as in a saturated observed data law). Those parameters are as follows: $3$ parameters for $p(X_1, X_2),$ $1$ parameter for $\text{OR}(R_1 = 0, R_2 = 0 | X_1, X_2) = f(R_1, R_2)$ (since the OR evaluated  at other levels of $R_1$ and $R_2,$ i.e., the reference values, is always one), $2$ parameters for $p(R_1  = 1 | R_2=1, X_2)$, and $2$ parameters for $p(R_2 =1 | R_1 = 1, X_1.)$ 

Examples of the three class of missing data  models that we are interested in are provided in Fig.~\ref{fig:seq-mar}(a), \ref{fig:seq-mnar_2}(a), and \ref{fig:seq-mnar_2}(d), where $X = \{X_1, X_2\}$. Here, we compare the full law parameterization of each example against the pattern-mixture parameterization as an illustrative step to show that the conditional independence restrictions on the full law impose restrictions on the observed data law. Given the MAR model in Fig.~\ref{fig:seq-mar}(a), the full law factorizes as $p(X_1, X_2) \times p(R_1) \times p(R_2 \mid R_1, X^*_1).$  Given the MNAR model in Fig.~\ref{fig:seq-mnar_2}(a) (without the dashed edge), the full law factorizes as $p(X_1, X_2) \times p(R_1 \mid X_2) \times p(R_2 \mid R_1).$ Given the MNAR model in Fig~\ref{fig:seq-mnar_2}(d), the full law factorizes as $p(X_1, X_2) \times p(R_1 \mid X_2) \times p(R_2 \mid X_1).$ In all the three examples, the full law requires $7$ parameters to encode the independencies (less than the number of parameters in the saturated observed data law). The above implies that there must be a testable implication, at least in the binary case, on the observed data laws of the three classes of missing data models that we consider. The parameter counting argument can be simply generalized to discrete data. Results in the main draft confirm that this  generalizes to situations where no distributional assumptions are made.

\subsection{On edges from proxy variables to missingness indicators} 
\label{app:proxy_edges}

The convention in previous work on missing data DAGs (e.g., \cite{mohan2013missing} and \cite{mohan2021graphical}) has often been to avoid including edges from proxy variables to missingness indicators. However, allowing for $X^*_i \rightarrow R_j$ edges  enables exploration of a broader class of missing data DAG models and MNAR mechanisms. For instance, the permutation MNAR model introduced by \cite{robins97non-a} can only be represented graphically if we permit proxy variables to point to missingness indicators. Without such edges, this model would lack a graphical characterization. A more comprehensive discussion on this topic can be found in \citep{nabi2022causal}. Models like the permutation model are particularly interesting as they represent nonparametrically saturated models with nonparametrically identified full laws. Thus, incorporating these edges allows our work to have a broader scope and naturally builds upon the foundations laid out in earlier research on testability in missing data DAGs, including the framework proposed by \cite{mohan2014testability}.

Here, we explore the substantive distinctions between models with edges $X^*_i \rightarrow R_j$ (as in the permutation model) and models with edges $X_i \rightarrow R_j$. To illustrate the dissimilarities between these two models, let us assume that $X_i$ is a binary variable, and we consider two structures: (1) $R_i \rightarrow R_j \leftarrow X_i$ and (2) $R_i \rightarrow R_j \leftarrow X^*_i$.
In the first structure, $p(R_j = 1 \mid R_i, X_i)$ has four parameters, with each parameter corresponding to a specific combination of values for $X_i$ and $R_i$. On the other hand, in the second structure, $p(R_j = 1 \mid R_i, X^*_i)$ only has three parameters due to the deterministic relationship between $R_i$ and $X^*_i$. These structural differences indicate qualitative differences as well. An $X_i \rightarrow R_j$ edge implies that the missing variable $X_i$ might have an impact on $R_j$. Conversely, an $X_i^* \rightarrow R_j$ edge suggests that the variable affects $R_j$ when it is observed, but when it is missing, its absence influences future missingness rather than its actual unobserved value.
These differences have implications for identification. If we change the edges in Fig.~3(a) to be $X_i \rightarrow R_j$, neither the full law nor the target law is identifiable. However, if we retain the edges as they are, the models are identifiable, as they represent the permutation model. Identifiability also plays a crucial role in determining testability, as discussed in the main manuscript.

Finally we  note that testing the absence of dashed edges involving proxy variables in Fig~3(a) is not entirely equivalent to testing edges involving their counterfactual counterparts. In other words, if for instance $R_2 \ci X_1 | R_1 = 1$ or equivalently $R_2 \ci X^*_1 | R_1 = 1$  holds in the observed data, there is no guarantee that $R_2$ and counterfactual $X_1$ are independent in the full law; because for the the independence in the full law to hold, we must show that $R_2 \ci X_1$ even among rows where $R_1 = 0$. This may be possible under a further assumption like \textit{faithful observability} used by \cite{tu2019causal} (which is a stronger assumption than standard faithfulness) where independences in the observed data ``do not lie'' about independences in the full data. But in the case where the full/target law is not identified, an assumption like this could be misleading – in this case $p(R_2 | R_1=0, X_1)$ is not identified and there is no way to confirm the validity of the test in the full data law. However, this was not a particular issue for the method proposed in \citep{tu2019causal}, as they consider a subclass of MNAR models where the full law is always identified. In future research, it would be interesting to explore  the additional constraints imposed by assumptions like faithful observability, which may lead to $X_i \rightarrow R_j$ edges resembling edges from a proxy variable rather than the actual underlying counterfactual.

\section{More on goodness-of-fit tests in the sequential MNAR model}
\label{app:likelihood-tests}

\subsection{General algorithm for goodness-of-fit tests  using likelihood approaches }

Algorithm~\ref{alg:seq-mnar} illustrates how to perform a sequential goodness-of-fit tests based on weighted likelihood-ratios for $K$ greater than $3$ variable in sequential MNAR models. 

\begin{algorithm}[!h]
	\caption{\textproc{Testing sequential MNAR} {\small $(\prec, \mathcal{M}, \mathcal{D}_n)$}}  \label{alg:seq-mnar}
	\begin{algorithmic}[1] 
		
		\State Let $\prec$ index variables by $k = 1, \ldots, K.$
		
		\vspace{0.2em}
		\State Let $\Omega_{K+1} = 1.$
		
		\vspace{0.2em}
		\For{$k \in \{K, \ldots, 2\}$}
		
		\vspace{0.2em}
		\State Let {\small $W_k(\beta^o_k) \coloneqq p(R_{k} | R_{\prec k}, X_{\succ k}; \beta^o_k)$} and 
		
		\hspace{0.4cm} {\small $W_k(\beta^a_k) \coloneqq p(R_k | R_{\prec k}, X_{\succ k}, X^*_{\prec k}; \beta^a_k)$}.  
		
		\State Estimate $\beta^o_k$ and $\beta^a_k$ via the following:
		{\small
			\begin{align*}
				\mathbb{P}_n \big[ \Omega_{k+1} \times U(\beta^o_k)  \big] = 0, \quad \mathbb{P}_n \big[ \Omega_{k+1} \times U(\beta^a_k)  \big] = 0,
			\end{align*}
		}%
		where $\mathbb{P}_n\big[ U(\beta^o_k) \big] = 0$ and $\mathbb{P}_n\big[ U(\beta^a_k) \big] = 0$ are estimating equations for $\beta^o_k$ and $\beta^a_k$ wrt the full law. 
		
		\vspace{0.2em}
		\State Compute a weighted likelihood-ratio as follows: 
		{\small
			\begin{align*}
				\rho = n\mathbb{P}_n \bigg[ \Omega_{k+1} \times \log\Big(  \frac{W_k(\widehat{\beta}^a_k)}{W_k(\widehat{\beta}^o_k)}   \Big)  \bigg]. 
			\end{align*}
		}%
		
		\vspace{-0.1cm}
		\State Test $\rho$ with $\alpha$ significance level. 
		
		\vspace{0.1cm}
		\If{$\mathcal{M}_o$ is rejected {\small (i.e., $R_k \not\ci X^*_{\prec k} | R_{\prec k}, X_{\succ k}$)}}
		\State \textbf{return} not sequential MNAR 
		
		\vspace{0.1cm}
		\Else{ $\Omega_{k+1} = \frac{\mathbb{I}(R_{\succ k} = 1)}{\prod_{j \succ k}^{K} W_j(\widehat{\beta}^o_{j}) }$. }
		\EndIf
		
		\EndFor
		\State \textbf{return} sequential MNAR
	\end{algorithmic}
\end{algorithm}

\subsection{Alternative supermodels in the sequential MNAR model}

Consider the m-DAG in Fig.~\ref{fig:seq-mnar_2}(a). We are interested in the absence of an edge between $X_1$ and $R_2$ which implies $R_2 \ci X_1 | R_1$. The no self-censoring supermodel is drawn in Fig.~\ref{fig:seq-mnar_2}(b) (with $R_1, R_2$ edge undirected). We can evaluate this independence by showing $p(R_2 | R_1, X_1)$ is not a function of $X_1$. See Appendix~B.2 for  details on how to set up such a test. 

For this, we use the following odds-ratio factorization of $p(R | X)$ \citep{chen2007semiparametric}:
{\small
	\begin{align}
		p(R \mid X) &= \frac{1}{Z(X)} \times p(R_1 \mid R_2 = 1, X_2)  
		\times p(R_2 \mid R_1 = 1, X_1) \times  \text{OR}(R_1, R_2 | X),  \label{eq:odds_two} 
	\end{align}	
}%
where $Z(X)$ is a normalizing term and $\text{OR}(R_1, R_2 | X)$ is the conditional odds ratio between $R_1$ and $R_2$. Since the no self-censoring model is identified, each piece above must be a function of observed data. This is trivial for the univariate conditionals, however, it can also be shown that $\text{OR}(R_1, R_2 | X) = f(R_1, R_2),$ i.e., is not a function of $X$ (see Appendix~A, Eq.~2.) By definition $p(R_2 | R_1, X_1) = p(R | X)/\sum_{R_2} p(R | X);$ to show $p(R_2 | R_1, X_1)$ is not a function of $X_1$, it suffices to show $p(R | X)$ is not a function of $X_1$ which using (\ref{eq:odds_two}) only requires us to show $p(R_2 | R_1=1, X_1)$ is not a function of $X_1$ which is easy to evaluate. This can be generalized to $K > 2$, but it involves higher order interactions terms in the odds-ratio parameterization, which is why we prefer the permutation model as our supermodel choice; see Appendix~C.1 for more details.

\section{More on goodness-of-fit tests with odds ratios}
\label{app:odds_test}

\subsection{Sequential MNAR model as a submodel of no self-censoring model}
\label{app:seq-mnar_noself}

\begin{figure}[!h] 
	\begin{center}
		\scalebox{0.6}{
			\begin{tikzpicture}[>=stealth, node distance=1.6cm]
				\tikzstyle{format} = [thick, circle, minimum size=1.0mm, inner sep=3pt]
				\tikzstyle{square} = [draw, thick, minimum size=4.5mm, inner sep=3pt]
				
				\begin{scope}[xshift=0.cm]
					\path[->, thick]
					node[format] (x11) {$X_1$}
					node[format, right of=x11, xshift=0.55cm] (x21) {$X_2$}
					node[format, right of=x21, xshift=0.55cm] (x31) {$X_3$}
					node[format, below of=x11] (r1) {$R_1$}
					node[format, below of=x21] (r2) {$R_2$}
					node[format, below of=x31] (r3) {$R_3$}
					node[format, below of=r1, yshift=-0.25cm] (x1) {$X^*_1$}
					node[format, below of=r2, yshift=-0.25cm] (x2) {$X^*_2$}
					node[format, below of=r3, yshift=-0.25cm] (x3) {$X^*_3$}
					
					(x11) edge[blue] (x21) 
					(r1) edge[blue] (r2)
					
					(x31) edge[blue] (r2)
					(x31) edge[blue] (r1)
					(x21) edge[blue] (r1)
					
					(x21) edge[blue] (x31) 
					(r2) edge[blue] (r3)
					
					(x11) edge[blue, bend left] (x31) 
					(r1) edge[blue, bend right=25] (r3)
					
					(x11) edge[gray, bend right=25] (x1)
					(x21) edge[gray, bend left=25] (x2)
					(x31) edge[gray, bend left=25] (x3)
					(r1) edge[gray] (x1)
					(r2) edge[gray] (x2)
					(r3) edge[gray] (x3)
					
					node[format, below of=x2, xshift=0cm, yshift=0.65cm] (a) {(a)}; 
				\end{scope}
				
				\begin{scope}[xshift=8cm]
					\path[->, thick]
					node[format] (x11) {$X_1$}
					node[format, right of=x11, xshift=0.55cm] (x21) {$X_2$}
					node[format, right of=x21, xshift=0.55cm] (x31) {$X_3$}
					node[format, below of=x11] (r1) {$R_1$}
					node[format, below of=x21] (r2) {$R_2$}
					node[format, below of=x31] (r3) {$R_3$}
					node[format, below of=r1, yshift=-0.25cm] (x1) {$X^*_1$}
					node[format, below of=r2, yshift=-0.25cm] (x2) {$X^*_2$}
					node[format, below of=r3, yshift=-0.25cm] (x3) {$X^*_3$}
					
					(x11) edge[blue] (x21) 
					(r1) edge[blue, -] (r2)
					(x11) edge[blue] (r2)
					
					(x31) edge[blue] (r2)
					(x31) edge[blue] (r1)
					(x21) edge[blue] (r1)
					
					(x21) edge[blue] (x31) 
					(r2) edge[blue, -] (r3)
					(x21) edge[blue] (r3)
					
					(x11) edge[blue, bend left] (x31) 
					(r1) edge[blue, bend right=25, -] (r3)
					(x11) edge[blue] (r3)
					
					(x11) edge[gray, bend right=25] (x1)
					(x21) edge[gray, bend left=25] (x2)
					(x31) edge[gray, bend left=25] (x3)
					(r1) edge[gray] (x1)
					(r2) edge[gray] (x2)
					(r3) edge[gray] (x3)
					
					node[format, below of=x2, xshift=0cm, yshift=0.65cm] (b) {(b)}; 
				\end{scope}
				
			\end{tikzpicture} 
		}
		
		\caption{ (a) Example of a sequential MNAR model; (b) The permutation supermodel.} 
		\label{app:fig:seq-mnar}
		\vspace{-0.5cm}
	\end{center}
\end{figure}
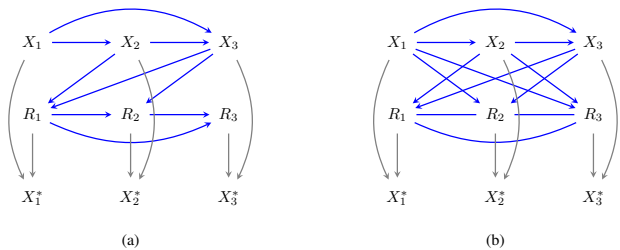

As mentioned in Remark 1, the sequential MNAR model can be viewed as a submodel of the no self-censoring model. This provides  a way to test independence restrictions of the form $R_k \ci X_{\prec k} \mid R_{-k}, X_{\succ k}.$ We provided an example with two variables using the m-DAG in Fig.~\ref{fig:seq-mnar_2}(a) and showed how to use odds-ratio parameterization of the missingness mechanism to test the absence of an edge between $X_1$ and $R_2$ which implied $R_2 \ci X_1 | R_1$. Extending the idea to sequential MNAR models with $K > 2$ involves higher order interaction terms in the odds-ratio parameterization. We use the sequential MNAR model with three variables, shown in Fig.~\ref{app:fig:seq-mnar}(a), to illustrate this point. The no self-censoring supermodel is shown in Fig.~\ref{app:fig:seq-mnar}(b). We are interested in testing the absence of $X_1 \rightarrow R_2, X_1 \rightarrow R_3, X_2 \rightarrow R_3$ edges which implies the independence restrictions: $R_3 \ci X_1, X_2 | R_1, R_2$ and $R_2 \ci X_1 | R_1, R_3, X_3.$ Let us focus on the former independence, i.e, $R_3 \ci X_1, X_2 | R_1, R_2$ which entails showing that $p(R_3 | R_1, R_2, X_1, X_2)$ is not a function of $X_1$ and $X_2.$ Note that $p(R_3 | R_1, R_2, X_1, X_2) = p(R | X) / \sum_{R_3} p(R | X).$ The odds-ratio parameterization of $p(R | X)$ is as follows: 

\vspace{-0.5cm}
{\small
	\begin{align*}
		p(R \mid X) 
		&= \frac{1}{Z} \times p(R_1 | R_2=R_3=1, X) \times p(R_2 | R_1=R_3=1, X) \times p(R_3 | R_1=R_2=1, X) \\
		&\hspace{1cm} \times \text{OR}(R_2, R_1 | R_3=1, X_1, X_2, X_3) \times \text{OR}(R_3, R_1, R_2 | X)  \\[0.2em]
		&= p(R_1 | R_2=R_3=1, X_2, X_3) \times p(R_2 | R_1=R_3=1, X_1, X_2) \times  p(R_3 | R_1=R_2=1, X_1, X_2) \\
		&\hspace{1cm} \times f(R_2, R_1, X_3) \times \text{OR}(R_3, R_1, R_2 | X). 
	\end{align*}
}%
The equality uses assumptions in the no self-censoring supermodel: $R_k \ci X_{k} | R_{-k}, X_{-k}, \forall k$ and the symmetry of the odds ratio to show $\text{OR}(R_2, R_1 | R_3=1, X_1, X_2, X_3) =f(R_1, R_1, X_3).$ Thus, to show $p(R_3 | R_1, R_2, X_1, X_2)$ is not a function of $X_1$ and $X_2$, it suffices to show that $p(R_3 | R_1=1, R_2=1, X_1, X_2) \times \text{OR}(R_3, R_1, R_2 | X)$ is not a function of $X_1, X_2.$ Here, we see the higher order interaction term $ \text{OR}(R_3, R_1, R_2 | X)$ appearing. Even though estimating equations have been discussed in \cite{malinsky2021semiparametric} to estimate these higher order terms, they make the tests more challenging. 

The above representation becomes more complex as the number of variables increase. This makes it clear why using the saturated permutation model is relatively easier to test the sequential MNAR models.

\subsection{Sequential MAR model as a submodel of permutation model} 
\label{app:seq-mar}

Here, we discuss odds ratio independence test as an alternative to likelihood-ratio goodness-of-fit test in sequential MAR models (as submodels of permutation model). The independence restrictions we would like to test are: $R_k \ci X_{\succ k} | R_{\prec k}, X^*_{\prec k}, \forall k$. We break down the independencies involving $R_k$ into $K - k$ individual tests, i.e., we would like to test $R_k \ci X_j | R_{\prec k}, X^*_{\prec k}, X_{\succ k, \prec j}, \forall X_j \in X_{\succ k}$, where $X_{\succ k, \prec j}$ denotes $\{X_{k+1}, \ldots, X_{j-1}\}$.  As mentioned in the main draft, the conditional independence $A \ci B | C$ holds if and only if $\text{OR}(A, B | C) = 1$ for all values of $A, B, C.$ Therefore, to show the independence between $R_k$ and $X_j$, we need to show that the following odds ratio is one for all levels of $R_k, X_j$ with statistical significance-level $\alpha$: 
{\small
	\begin{align*}
		&\text{OR}(R_k = r_k, X_j=x_j \mid R_{\prec k}, X^*_{\prec k}, X_{\succ k, \prec j}) \\ 
		&\hspace{1cm} = \frac{p(R_k  = r_k \mid X_j = x_j, R_{\prec k}, X^*_{\prec k}, X_{\succ k, \prec j}; {\beta}^a_k)}{p(R_k = 1 \mid X_j = x_j, R_{\prec k}, X^*_{\prec k}, X_{\succ k, \prec j}; {\beta}^a_k)} 
		\times 
		\frac{p(R_k  = 1 \mid X_j = 1, R_{\prec k}, X^*_{\prec k}, X_{\succ k, \prec j}; {\beta}^a_k)}{p(R_k = r_k \mid X_j = 1, R_{\prec k}, X^*_{\prec k}, X_{\succ k, \prec j}; {\beta}^a_k)}.
	\end{align*}
}%
To estimate the odds ratio, we need an estimate of $\beta^a_k$ parameters. We use weighted estimating equations to estimate $\beta^a_k$. The intuition is as follows. Given that we have the permutation model as the supermodel, the independence restriction involving $R_k$ and $X_j$ is equivalent to the following Verma constraint: 
\begin{align*}
	R_k \ci X_j \mid R_{\prec k}, X^*_{\prec k}, X_{\succ k, \prec j}, \blue{\text{do}(R_{\succ k, \prec j+1} = 1)}, \  \forall X_j \in X_{\succ k}, 
\end{align*}
where the post intervention distribution is defined as follows: 
\begin{align*}
	p(. \mid \text{do}(R_{\succ k, \prec j+1} = 1)) = \displaystyle  \frac{p(V)}{ \prod_{i=k+1}^j \ p(R_i \mid \pa_{\cal G}(R_i)) }\Bigg|_{R_{\succ k, \prec j+1} = 1}. 
\end{align*}
Let $W_k(\beta_k) \coloneqq p(R_k | R_{\prec k}, X^*_{\prec k}, X_{\succ k, \prec j}, X_j; \beta_k)$ and let $\mathbb{P}_n\big[ U(\beta_k) \big] = 0$ be an unbiased estimating equation for $\beta_k$ wrt the full law (i.e., had there been no missingness). We can estimate $\beta_k$ via the following weighted estimating equation:
\begin{align*}
	\mathbb{P}_n \bigg[  \frac{\mathbb{I}(R_{\succ k, \prec j+1} = 1)}{\prod_{i=k+1}^j \omega_i(\widehat{\eta}_i)}  \times U(\beta_k)  \bigg] = 0, 
\end{align*}
where $\omega_i(\eta_i) \coloneqq p(R_i \mid \pa_{\cal G}(R_i); \eta)$, and $\widehat{\eta}_i$ denotes an estimate of $\eta_i$. 

Since we have to evaluate the odds ratio for all values of $X_j$, the tests can become expensive in discrete cases and even more challenging in continuous cases, \citep{chen2021semiparametric}. Hence, the likelihood-ratio test in Algorithm~\ref{alg:seq-mar} might be preferred over odds ratio independence tests for larger graphs. 

\subsection{Sequential MNAR model as a submodel of permutation model}
\label{app:alg_seq-mnar_odds}

The independence restrictions we would like to test are: $R_k \ci X^*_{\prec k} | R_{\prec k}, X_{\succ k}, \forall k$. We break down the independencies involving $R_k$ into $k-1$ individual tests, i.e., $R_k \ci X^*_j | R_{\prec k}, X_{\succ k}, X^*_{\prec j}, \forall X^*_j \in X^*_{\prec k}$. As mentioned in the main draft, this is a context-specific independence restriction and is equivalent to $R_k \ci X_j | R_{\prec k} \setminus R_j, R_j = 1, X_{\succ k}, X^*_{\prec j}$. This independence holds if and only if the following odds ratio is one for all levels of $X_j$ with statistical significance-level $\alpha:$ 
{\small
	\begin{align*}
		&\text{OR}(R_k = r_k, X_j=x_j \mid R_{\prec k} \setminus R_j, R_j = 1, X_{\succ k}, X^*_{\prec j}) \\ 
		&\hspace{0.5cm} = \frac{p(R_k  = r_k \mid X_j = x_j, R_{\prec k} \setminus R_j, R_j = 1, X_{\succ k}, X^*_{\prec j}; {\beta}^a_k)}{p(R_k = 1 \mid X_j = x_j, R_{\prec k} \setminus R_j, R_j = 1, X_{\succ k}, X^*_{\prec j}; {\beta}^a_k)} 
		\times 
		\frac{p(R_k  = 1 \mid X_j = 1, R_{\prec k} \setminus R_j, R_j = 1, X_{\succ k}, X^*_{\prec j}; {\beta}^a_k)}{p(R_k = r_k \mid X_j = 1, R_{\prec k} \setminus R_j, R_j = 1, X_{\succ k}, X^*_{\prec j}; {\beta}^a_k)}.
	\end{align*}
}%
We can estimate the odds ratio by estimating the parameters $\beta^a_k$. We use weighted estimating equations to estimate the parameters and the intuition behind the choice of weights is that the restriction between $R_k$ and $X^*_j$ can be viewed as the following Verma constraint (under the permutation supermodel): 
\begin{align*}
	R_k \ci X^*_j | R_{\prec k}, X_{\succ k}, X^*_{\prec j},  \blue{\text{do}(R_{\succ k} = 1)}, \  \forall X^*_j \in X^*_{\prec k}. 
\end{align*}
Let $W_k(\beta^a_k) \coloneqq p(R_k | X_j, R_{\prec k} \setminus R_j, R_j = 1, X_{\succ k}, X^*_{\prec j}; \beta^a_k)$ and let $\mathbb{P}_n\big[ U(\beta^a_k) \big] = 0$ is unbiased estimating equation for $\beta^a_k$ wrt the full law (had there been no missingness). We can estimate $\beta^a_k$ via the following weighted estimating equation:
\begin{align*}
	\mathbb{P}_n \bigg[ \displaystyle  \frac{\mathbb{I}(R_{\succ k} = 1)}{\prod_{j=k+1}^{K} p(R_j | \pa_{\cal G}(R_j); \widehat{\eta}_j) }  \times U(\beta^a_k)  \bigg] = 0,
\end{align*}
where  $\widehat{\eta}_j$ is an estimate of $\eta_j$ that parameterize the conditional density of $p(R_j | \pa_{\cal G}(R_j))$. 

Similar to the sequential MAR model, the goodness-of-fit test based on odds ratio independence test can be rather challenging with continuous variables. Hence, the weighted likelihood-ratio tests might still be preferred.

\subsection{Block-parallel model as a submodel of no self-censoring}
\label{app:alg_par-odds}

\begin{algorithm}[!h]
	\caption{\textproc{Testing block-parallel} {\small $(\mathcal{M}, \mathcal{D}_n)$}}  \label{alg:block-par}
	\begin{algorithmic}[1] 
		
		\vspace{0.2em}
		\For{$k \in \{1, \ldots, K-1\}$} 
		
		\vspace{0.2em}
		\State Let $W_k(\beta_k) \coloneqq p(R_k = 1 \mid R_{-k}=1, X_{-k}; \beta_k)$. 
		
		\vspace{0.2em}
		\State Estimate $\beta_k$ (denoted by $\widehat{\beta}_k$). 
		
		\EndFor
		
		\vspace{0.2em}
		\For{ each pair $k, j \in \{1, \dots, K\}$ s.t. $k\not=j$}
		
		\vspace{0.35em}
		\State Let {\small $\theta(r_k, r_j) = \text{OR}(R_k=R_j=0 \mid R_{-kj}=1, X) \!\!\!$}
		
		\vspace{0.35em}
		\State Compute $\theta(R_k = 0, R_j = 0)$ via the following:
		{\scriptsize
			\begin{align*}
				\frac{\mathbb{P}_n \Big[ \prod_{i \not= \{k,j\}} R_i \times (1-R_k) \times (1-R_j)  \Big] }{
					\mathbb{P}_n \bigg[  \prod_{i = 1}^K R_i \times \displaystyle \frac{(1-W_k(\widehat{\beta}_k)) \times (1- W_j(\widehat{\beta}_j))}{W_k(\widehat{\beta}_k) \times W_j(\widehat{\beta}_j)}  \bigg] }
			\end{align*}
		}
		
		\State Test $\theta(R_k \!= \!0,  \! R_j \! = \! 0) \! = \! 1$ at significance level $\alpha$ 
		
		\vspace{0.15cm}
		\If{test fails {\small (i.e., $R_k \not\ci R_j | X$)}}
		\vspace{0.1cm}
		\State \textbf{return} not block-parallel MNAR 
		\EndIf
		
		\vspace{0.1cm}
		\EndFor
		\State \textbf{return} block-parallel MNAR
	\end{algorithmic}
\end{algorithm}

\section{Proofs}
\label{app:proof}

{\large \bf Theorem~\ref{thm:seq-mar}.} 
%
The intervention distribution $p(X, R \setminus R_{\succ k}, X^* | \text{do}(R_{\succ k} =1))$ factorizes wrt a CDAG ${\cal G}^*$ where  edges into $R_{\succ k}$ have been removed from the sequential MAR graph ${\cal G}.$ Factorization of this intervention distribution wrt a CDAG preserves the global Markov property, i.e., d-separation can be used to read dormant independencies in the intervention distribution. In ${\cal G}^*$ we have $R_k \ci X_{\succ k} | R_{\prec k}, X^*_{\prec k}$ by d-separation implying the same independence holds in the intervention distribution. Finally, testability of this dormant independence from observed data follows from the fact that the propensity scores $p(R_j | \pa_{\cal G}(R_j))$ for each $R_j \in R_{\succ k}$ is identified under the restrictions implied by the graph ${\cal G}$ (identification is trivial since the sequential MAR model is a submodel of a permutation model that is fully identified), and upon intervention to $R_{\succ k}=1,$ each previously partially observed variable $X_j \in X_{\succ k}$ is now observed via a consistency argument $X_j = X_j^*.$

\vspace{0.5cm}
{\large \bf Theorem~\ref{thm:seq-mnar}.} 
%
The proof is very similar to the proof of Theorem~\ref{thm:seq-mar}. Interventions on $R_{\succ k}$ preserve the global Markov property and  propensity scores of $R_{\succ k}$ are all identified as functions of observed data (since sequential MNAR is a submodel of fully identified permutation model). The m-CDAG we obtain after intervening on $R_{\succ k}$ and setting them to $1$ is a graph where all incoming edges into $R_{\succ k}$ are removed and all $X_{\succ k}$ are observed random variables. Thus the dormant independence are direct functions of observed data. 

\vspace{0.5cm}
{\large \bf Theorem~\ref{thm:block-par}.} 
%
Given the restrictions of a block-parallel model, we note that including $R_{-kj}$ in the conditioning set of independence $R_k \ci R_j |  X$ does not spoil the independence.  Hence, we can equivalently look at $R_k \ci R_j |  X, R_{-kj} = 1$. Further, we know this independence holds if and only if $\text{OR}(R_k,  R_j |  X, R_{-kj} = 1) = 1.$ All we need to show now is that $\text{OR}(R_k,  R_j |  X, R_{-kj} = 1)  = \text{OR}(R_k,  R_j |  X_{-kj}, R_{-kj} = 1).$ Using an odds-ratio parameterization of $p(R_k, R_j | X, R_{-kj} = 1)$ we have: 
{\small
	\begin{align*}
		\text{OR}(R_k = r_k, R_j=r_j \mid X, R_{-kj} = 1)
		& = \frac{p(R_k=r_k \mid R_j=r_j, X, R_{-kj} = 1)}{p(R_k = 1 \mid R_j=r_j, X, R_{-kj} = 1)} \times \frac{p(R_k  = 1 \mid R_j = 1, X, R_{-kj} = 1)}{p(R_k=r_k \mid R_j = 1, X, R_{-kj} = 1)} \\ 
		& = \frac{p(R_k=r_k \mid R_j=r_j, X_{-k}, R_{-kj} = 1)}{p(R_k = 1 \mid R_j=r_j, X_{-k}, R_{-kj} = 1)} \times \frac{p(R_k  = 1 \mid R_j = 1, X_{-k}, R_{-kj} = 1)}{p(R_k=r_k \mid R_j = 1, X_{-k}, R_{-kj} = 1)} \\ 
		&= f_1(R_k, R_j, X_{-k}, R_{-kj} = 1).
	\end{align*}
}%
The second equality holds because $R_k \ci X_k | R_{-k}, X_{-k}$, and 
{\small
	\begin{align*}
		\text{OR}(R_j=r_j, R_k = r_k \mid X, R_{-kj} = 1)
		& = \frac{p(R_j=r_j \mid R_k=r_k, X, R_{-kj} = 1)}{p(R_j = 1 \mid R_k=r_k, X, R_{-kj} = 1)} \times \frac{p(R_j = 1 \mid R_k = 1, X, R_{-kj} = 1)}{p(R_j=r_j \mid R_k = 1, X, R_{-kj} = 1)} \\ 
		& = \frac{p(R_j=r_j \mid R_k=r_k, X_{-j}, R_{-kj} = 1)}{p(R_j = 1 \mid R_k=r_k, X_{-j}, R_{-kj} = 1)} \times \frac{p(R_j  = 1 \mid R_k = 1, X_{-j}, R_{-kj} = 1)}{p(R_j=r_j \mid R_k = 1, X_{-j}, R_{-kj} = 1)} \\ 
		&= f_2(R_k, R_j, X_{-j}, R_{-kj} = 1).
	\end{align*}
}
The second equality holds because $R_j \ci X_j | R_{-j}, X_{-j}$. Due to symmetry of odds ratio, $f_1(R_k, R_j, X_{-k}, R_{-kj} = 1)$ and $f_2(R_k, R_j, X_{-j}, R_{-kj} = 1)$ must be equal. This implies $\text{OR}(R_k,  R_j |  X, R_{-kj} = 1)  = \text{OR}(R_k,  R_j |  X_{-kj}, R_{-kj} = 1)$ (all a function of observed data).  

Even though the odds ratio is a function of observed data, estimation of odds ratio is not straightforward. We rely on the estimating equations discussed in this Appendix and \cite{malinsky2021semiparametric} to estimate the odds ratios. 

\vspace{0.5cm}
{\large \bf Theorem \ref{thm:criss-cross}}. To prove this result, it suffices to show that the target law in the  criss-cross structure on two variables (drawn on the right hand side) is not non-parametrically identified. For this purpose, we provide an example of two different full laws that factorize according to the criss-cross model, but map into the same observed data law. 
\\

\begin{minipage}{0.8\textwidth}
	\scalebox{0.7}{
		
		\begin{tabular}{ c | c }
			$X_1$ & $p(X_1)$ \\ \hline
			$0$  & $\red{a}$     \\ 
			$1$  & $\red{1-a}$ 
		\end{tabular}
		
		\hspace{0.5cm}
		
		\begin{tabular}{ c  c | c }
			$X_2$ & $X_1$ &  $p(X_2 \mid X_1)$  \\ \hline
			$0$  & $0$ & $\red{b}$     \\ 
			$1$  & $0$ & $\red{1-b}$ \\  \hline 
			$0$  & $1$ & $\red{c}$     \\ 
			$1$  & $1$ & $\red{1-c}$ 
		\end{tabular}
		
		\hspace{0.5cm}
		
		\begin{tabular}{ c  c | c }
			$R_1$ & $X_2$ &  $p(R_1 \mid X_2)$  \\ \hline
			$0$  & $0$ & $\red{d}$     \\ 
			$1$  & $0$ & $\red{1-d}$ \\  \hline 
			$0$  & $1$ & $\red{e}$     \\ 
			$1$  & $1$ & $\red{1-e}$ 
		\end{tabular}
		
		\hspace{0.5cm}
		
		\begin{tabular}{ c  c c | c}
			$R_2$ & $R_1$ & $X_1$ & $p(R_2 \mid R_1, X_1)$   \\ \hline
			$0$  & $0$ & $0$  & $\red{f}$     \\
			$1$  & $0$ & $0$ & $\red{1 - f}$ \\ \hline 
			$0$  & $0$ & $1$ & $\red{g}$     \\
			$1$  & $0$ & $1$ & $\red{1- g}$  \\  \hline 
			$0$  & $1$ & $0$  & $h$   \\
			$1$  & $1$ & $0$ & $1-h$ \\ \hline 
			$0$  & $1$ & $1$ & $i$     \\
			$1$  & $1$ & $1$ & $1- i$ 
		\end{tabular}
	}
\end{minipage}
\begin{minipage}{0.15\textwidth}
	\begin{center}
		\scalebox{0.8}{
			\begin{tikzpicture}[>=stealth, node distance=1.5cm]
				\tikzstyle{format} = [thick, circle, minimum size=1.0mm,
				inner sep=0pt]
				\begin{scope}
					\path[->, thick]
					node[format] (x11) {$X_1$}
					node[format, right of=x11] (x21) {$X_2$}
					node[format, below of=x11] (r1) {$R_1$}		
					node[format, below of=x21] (r2) {$R_2$}
					node[format, below of=r1] (x1) {$X^*_1$}
					node[format, below of=r2] (x2) {$X^*_2$}
					(x11) edge[blue] (x21)
					(x21) edge[blue] (r1)
					(x11) edge[blue] (r2)
					(r1) edge[blue] (r2)
					(r1) edge[gray] (x1)
					(r2) edge[gray] (x2)
					(x11) edge[gray, bend right] (x1)
					(x21) edge[gray, bend left] (x2)
					;
				\end{scope}
			\end{tikzpicture}
		}
	\end{center}
\end{minipage}
%
\begin{table}[h]
	\scalebox{0.8}{
		\begin{tabular}{ | c  c | c  c | c  | c  c | c | }
			\hline
			$R_1$  & $R_2$   & $X_1$   &  $X_2$   & \blue{p(FULL LAW)}  & $X^*_1$   & $X^*_2$   &  \blue{p(OBSERVED LAW)}    \\ \hline
			\multirow{4}{*}{0} & \multirow{4}{*}{0} & 0     & 0    &  $abdf$   & \multirow{4}{*}{?} & \multirow{4}{*}{?} & \multirow{4}{*}{$d\Big[ abf + (1-a)cg \Big] + e\Big[ a(1-b)f + (1-a)(1-c)g \Big]$}   \\ 
			&   &  1 & 0  & $(1-a)cdg$   &   &  &  \\
			&   &  0 & 1  & $a(1-b)ef$   &   &  &  \\
			&   &  1 & 1  & $(1-a)(1-c)eg$   &   &  &  \\  
			
			\hline \hline 
			
			\multirow{4}{*}{0} & \multirow{4}{*}{1} & 0     & 0    &  $abd(1-f)$   & \multirow{4}{*}{?} & \multirow{2}{*}{$0$}   & \multirow{2}{*}{$d\Big[ ab(1-f) + (1-a)c (1-g)\Big]$}   \\ 
			&   &  1 & 0  & $(1-a)cd(1-g)$   &  &  &  \\
			&   &  0 & 1  & $a(1-b)e(1-f)$   &  & \multirow{2}{*}{$1$}  &  \multirow{2}{*}{$e\Big[ a(1-b)(1-f)+ (1-a)(1-c)(1-g)\Big]$}  \\
			&   &  1 & 1  & $(1-a)(1-c)e(1-g)$   &  &  &  \\  
			
			\hline \hline 
			
			\multirow{4}{*}{1} & \multirow{4}{*}{0} & 0     & 0    &  $ab(1-d)h$   &  \multirow{2}{*}{$0$}   & \multirow{4}{*}{?} & \multirow{2}{*}{$ah\Big[ b(1-d) + (1-b)(1-e)\Big]$}   \\ 
			&   &  1 & 0  & $(1-a)c(1-d)i$   &  &  &  \\
			&   &  0 & 1  & $a(1-b)(1-e)h$   &  \multirow{2}{*}{$1$} &  &  \multirow{2}{*}{$(1-a)i\Big[ c(1-d)+ (1-c)(1-e)\Big]$}  \\
			&   &  1 & 1  & $(1-a)(1-c)(1-e)i$   &  &  &  \\  
			
			\hline \hline 
			
			\multirow{4}{*}{1} & \multirow{4}{*}{1} & 0     & 0    &  $ab(1-d)(1-h)$   &  $0$  & $0$  & $ab(1-d)(1-h)$   \\ 
			&   &  1 & 0  & $(1-a)c(1-d)(1-i)$   & $1$  &  $0$ &   $(1-a)c(1-d)(1-i)$  \\
			&   &  0 & 1  & $a(1-b)(1-e)(1-h)$   &  $0$ &  $1$ &  $a(1-b)(1-e)(1-h)$   \\
			&   &  1 & 1  & $(1-a)(1-c)(1-e)(1-i)$   &  $1$ &  $1$ &  $(1-a)(1-c)(1-e)(1-i)$  \\  \hline
			
		\end{tabular}
	}
\end{table}


A concrete example is as follows: \\
\begin{minipage}{0.7\textwidth}
	\scalebox{0.7}{
		
		\begin{tabular}{ c | c | c }
			\multirow{2}{*}{$X_1$} &  \multicolumn{2}{c}{$p(X_1)$}  \\ \cline{2-3}
			& $M_1$ & $M_2$ \\ \hline 
			$0$  & $7/15$  &  $5/11$  \\ 
			$1$  & $8/15$ & $6/11$ 
		\end{tabular}
		
		\hspace{1cm}
		\vspace{0.5cm}
		
		\begin{tabular}{ c  : c | c | c}
			\multirow{2}{*}{$X_2$} & \multirow{2}{*}{$X_1$} & \multicolumn{2}{c}{$p(X_2 \mid X_1)$}   \\ \cline{3-4}
			& & $M_1$ & $M_2$ \\ \hline
			$0$  & $0$  & $6/7$  & $4/5$  \\
			$1$  & $0$  & $1/7$ & $1/5$  \\ \hline 
			$0$  & $1$  & $3/4$  & $2/3$  \\
			$1$  & $1$  & $1/4$ & $1/3$ 
		\end{tabular}
		
		\hspace{0.5cm}
		\vspace{0.5cm}
		
		\begin{tabular}{ c : c | c | c}
			\multirow{2}{*}{$R_1$} & \multirow{2}{*}{$X_2$} & \multicolumn{2}{c}{$p(R_1 \mid X_2)$}   \\ \cline{3-4}
			& & $M_1$ & $M_2$ \\ \hline
			$0$  & $0$  & $19/20$  &  $189/200$  \\
			$1$  & $0$  & $1/20$ & $11/200$ \\ \hline 
			$0$  & $1$  & $85/100$  & $89/100$  \\
			$1$  & $1$  & $15/100$  & $11/100$
		\end{tabular}
		
		\hspace{1cm}
		\vspace{0.5cm}
		
		\begin{tabular}{ c : c c | c | c}
			\multirow{2}{*}{$R_2$} & \multirow{2}{*}{$R_1$} & \multirow{2}{*}{$X_1$} & \multicolumn{2}{c}{$p(R_2 |\mid R_1, X_1)$}   \\ \cline{4-5}
			& & & $M_1$ & $M_2$ \\ \hline
			$0$  & $0$ & $0$  & $268/323$  &  $7636/16821$  \\
			$1$  & $0$ & $0$  & $55/323$ & $9185/16821$ \\ \hline 
			$0$  & $0$ & $1$  & $208/323$   & $16216/16821$ \\
			$1$  & $0$ & $1$  & $115/323$  & $605/16821$ \\ \hline 
			$0$  & $1$ & $0$  & $1/2$ & $1/2$ \\
			$1$  & $1$ & $0$  & $1/2$ & $1/2$ \\ \hline 
			$0$  & $1$ & $1$  & $1/2$ & $1/2$ \\
			$1$  & $1$ & $1$  & $1/2$  & $1/2$
		\end{tabular}
	}
\end{minipage}
\vspace{0.25cm}
\begin{table}[h]
	\begin{center}
		\scalebox{0.85}{
			\begin{tabular}{ | c  c | c  c | c  : c  | c  c | c | }
				\hline
				\multirow{2}{*}{$R_1$}  & \multirow{2}{*}{$R_2$}  & \multirow{2}{*}{$X_1$} &  \multirow{2}{*}{$X_2$}  & \multicolumn{2}{c |}{$p(R, X)$} & \multirow{2}{*}{$X^*_1$} & \multirow{2}{*}{$X^*_2$} & $p(R, X^*)$   \\ \cline{5-6} \cline{9-9}
				
				& & &  & $M_1$ & $M_2$ & & & $M_1 = M_2$  \\ \hline
				
				\multirow{4}{*}{0} & \multirow{4}{*}{0} & 0     & 0    &  $134/425$ & $3818/24475$ & \multirow{4}{*}{?} & \multirow{4}{*}{?} & \multirow{4}{*}{$68/100$} \\ 
				&   &  1 & 0  & $104/425$  & $8108/24475$ &   &  &  \\
				&   &  0 & 1  & $67/1425$ & $1118/30439$ &   &  &\\
				&   &  1 & 1  & $104/1425$ & $8108/51975$ &   &  & \\   
				
				\hline 
				
				\multirow{4}{*}{0} & \multirow{4}{*}{1} & 0     & 0    &  $11/170$ & $167/890$& \multirow{4}{*}{?} & \multirow{2}{*}{$0$}   & \multirow{2}{*}{$2/10$}  \\ 
				&   &  1 & 0 & $23/170$  & $11/890$ &  &  &  \\
				&   &  0 & 1  &  $11/1140$ & $167/3780$ &  & \multirow{2}{*}{$1$}  &  \multirow{2}{*}{$1/20$} \\
				&   &  1 & 1  & $23/570$ & $11/1890$ &  &  & \\ 
				
				\hline 
				
				\multirow{4}{*}{1} & \multirow{4}{*}{0} & 0     & 0   &  $1/100$ & $1/100$  &  \multirow{2}{*}{$0$} & \multirow{4}{*}{?} & \multirow{2}{*}{$3/200$}  \\ 
				&   &  1 & 0  & $1/100$ & $1/100$ &  &  &  \\
				&   &  0 & 1 & $1/200$ & $1/200$ &  \multirow{2}{*}{$1$}  & &  \multirow{2}{*}{$2/100$}  \\
				&   &  1 & 1  & $1/100$ & $1/100$ &  &  & \\ 
				
				\hline  
				
				\multirow{4}{*}{1} & \multirow{4}{*}{1} & 0     & 0    &  $1/100$ & $1/100$ & $0$ & $0$  &  $1/100$  \\     
				&  & 1  & 0  &  $1/100$ & $1/100$  & $1$   & 0  &  $1/100$  \\       
				&  & 0   & 1   &  $1/200$ & $1/200$ & $0$   & 1  &  $1/200$  \\ 
				&  & 1   & 1   &  $1/100$ & $1/100$  & $1$   & 1  &  $1/100$  \\  \hline
			\end{tabular}
		}
	\end{center}
\end{table}

From the above example, we see that none of the parameters in red are identified.


\clearpage
\section{Simulations}
\label{app:sims} 

As mentioned in the main draft, we describe three sets of simulations to illustrate the key results and the utility of our proposed methods -- each set focuses on a class of missing data models that we considered in the main draft. For each simulation, we generate four random variables from either a multivariate normal distribution or  binomial distribution. We induce missing values in all four variables according to a missingness mechanism that follows restrictions of either sequential MAR, sequential MNAR, block-parallel, or supermodels of them. All code necessary to reproduce our simulations is included with this submission. The data generating mechanism is described as follows.  

\underline{Generating $X$}: For Gaussian data, we generate four random variables from multivariate normal distribution with mean zero and covariance matrix $\sigma$ where the $ij$-th entry is $\sigma_{ij} = 1- | i-j| \times 0.25.$  For binary data, variable $X_k$ is generated from a binomial distribution with the probability of observing $X_k=1$ given $X_{\prec k}$ equals to $\text{expit} \big(a^0_{x_k} + \sum_{j \prec k} a^j_{x_k} \times X_j )$, where $\text{expit}(x) = 1/(1+\exp(-x))$ and parameters $a^j_{x_k}$ (for all $k = 1, \ldots, K$ and $j \prec k$) are generated uniformly from the $(-1, 1)$ interval. 

\underline{Generating $R$}:  In each class of missing data model, we consider generating $R$ according to two scenarios: one where the restrictions in the  missing data model we would like to test hold true (the null hypothesis should be accepted)  and one where the restrictions are violated (the null hypothesis should be rejected in favor of accepting the corresponding supermodel). All missingness indicators are generated from binomial distributions. The details on missing data parameters are as follows. 
\begin{align}
	p(R_k = 1 \mid R_{\prec k}, X^*_{\prec k}, \blue{X_{\succ k}})  &=  \text{expit} \ \big(a^0_{k} + \sum_{j \prec k} b^j_{k} \times R_j +  c^{j}_{k} \times R_jX^*_j  \ \blue{ + \sum_{i \succ k} d^i_k \times X_i } \big),  \ k = 1, \ldots,  4 \quad \text{(Simulation 1)} 
	\nonumber 	 \\
	p(R_k = 1 \mid R_{\prec k}, X_{\succ k}, \blue{X^*_{\prec k}})  &=  \text{expit} \ \big(a^0_{k} +  \sum_{i \succ k} d^i_{k}  \times X_i  + \sum_{j \prec k} b^j_{k} \times R_j  \ \blue{+  \ c^j_k \times R_jX^*_j }   \big), \  k = 1, \ldots,  4 \quad \text{(Simulation 2)}  
	\nonumber	\\ 
	p(R_k = 1 \mid X_{-k}) &=  \text{expit} \ \big(a^0_{k} + \sum_{j \not= k} b^j_{k} \times X_j) ,  \ k = 1, \ldots,  4 \quad \text{(Simulation 3)}.  
	\label{eq:sims_R}
\end{align} 

Addition of the blue terms simulate scenarios where the independence assumptions we would like to test are violated. All the parameters are randomly generated from a uniform distribution.  In order to control the proportion of missing values, we run the experiments with three different ranges for the uniform distribution: $(-1, 1), (-0.5, 1.5),$ and $(0, 2).$ 

\underline{Generating $X^*$}:  For each given sample, if $R_k = 1$ then $X^*_k = X$, otherwise $X^* = \text{NA}$. 

Our objective is to test the missing data restrictions by relying only on observed data, i.e.,  $(R, X^*)$ samples. 

\vspace{0.25cm}
{\bf Simulation 1.}  In the first set of simulations, we focused on testing the sequential MAR model defined via the set of restrictions in (\ref{eq:seq-mar}). The results were provided and discussed in the main draft. 

We briefly add that when true underlying missingness mechanism satisfies the assumptions of the sequential MAR model, missingness indicators are generated from (\ref{eq:sims_R}) without the blue terms. When the restrictions are no longer valid, missingness indicators are generated from (\ref{eq:sims_R}) with the blue terms.

\vspace{0.25cm}
{\bf Simulation 2.} In the second set of simulations we focus on testing the sequential MNAR model defined via the set of restrictions in (\ref{eq:no-colluder}). We follow Algorithm~\ref{alg:seq-mnar} to test the independence restrictions, which entails running a total of $K-1$ tests. Our test statistic is $2\times\rho$ and we use a chi-square distribution with $k-1$ degrees of freedom to evaluate the goodness-of-fits --  the degree of freedom is chosen as the difference between number of parameters in $W_k(\beta^a_k)$ and $W_k(\beta^0_k)$, as defined in the algorithm. If the p-values are all greater than $0.05$, we accept the sequential MNAR model. 

\begin{figure}[t]
	\centering
	\begin{subfigure}{.5\textwidth}
		\centering
		\includegraphics[scale=0.31]{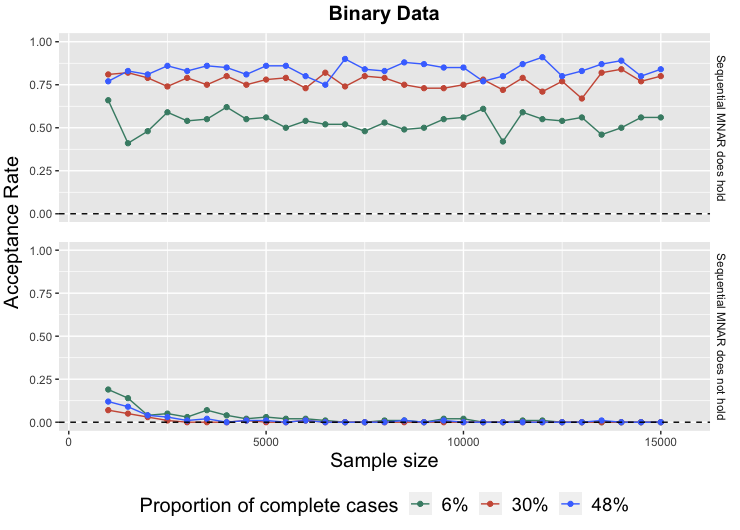}
		\label{fig:mnar_bin}
	\end{subfigure}%
	\begin{subfigure}{.5\textwidth}
		\centering
		\includegraphics[scale=0.31]{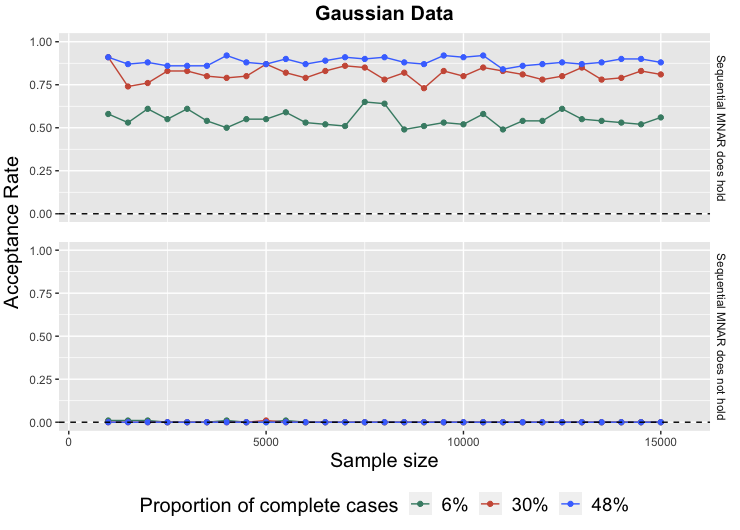}
		\label{fig:mnar_cont}
	\end{subfigure}
	\caption{Results on testing \textbf{sequential MNAR} models. In the top row, the sequential MNAR model captures the true underlying missingness mechanism. The assumptions of sequential MNAR model are violated in the bottom row. } 
	\label{fig:mnar_sim}
\end{figure}

For a fixed sample size, we simulate $100$ different datasets and calculate the acceptance rate of a sequential MNAR model. The acceptance rate is plotted as a function of sample size in Fig.~\ref{fig:mnar_sim}. The sample size ranges from $1,000$ to $15,000$ with $500$  increments. In each panel, there are three plots that vary in terms of the proportion of complete cases in the dataset, i.e, $6\%, 30\%, 48\%$. The top row illustrates the results when the true underlying missingness mechanism satisfies the assumptions of the sequential MNAR model  (missingness indicators are generated from (\ref{eq:sims_R}) without the blue terms) and the bottom row illustrates results for when the restrictions are no longer valid (missingness indicators are generated from (\ref{eq:sims_R}) with the blue terms).  As it is shown, the acceptance rate is quite low when the independence restrictions of a sequential MNAR model are not valid; even when we only have $6\%$ of complete cases the tests perform well. When the sequential MNAR model assumptions are true, the acceptance rate increases as missing rate decreases and reaches very close to $1$ when we have only $48\%$ complete cases.  


\vspace{0.25cm}
{\bf Simulation 3.}  In the third set of simulations we focus on testing independencies between missingness indicators in a block-parallel MNAR model defined via the set of restrictions in (\ref{eq:block-par}). Testing the full model requires following Algorithm~\ref{alg:block-par} which entails running a total of $\binom{K}{2}$ tests (between all distinct pairs of missingness indicators.) For illustration purposes, we focus on testing only one pair of missingness indicator in two different scenarios: one where the true underlying missingness mechanism follows the restrictions of a block-parallel model -- thus $R_k \in R$ is generated using (\ref{eq:sims_R}), and one where the missingness mechanism factorizes as $\prod_{k = 1}^K p(R_k | R_{\succ k}, X_{\prec k})$ which is still a submodel of the no-self censoring model but violates the assumptions of the block-parallel model. We focus on testing the independence $R_1 \ci R_2 | X$ by calculating the odds ratio $\theta \coloneqq \text{OR}(R_1 = 0, R_2=0 | X)$ via the following  estimating equation and showing that the value is one. 
{\small
	\begin{align*}
		&\mathbb{P}_n \Big[  R_1 \times R_2 \times R_3  \times \frac{p(R_1 = 0 \mid R_2=1, R_3=1, X_2, X_3 ) \times p(R_2 = 0 \mid R_1=1, R_3=1, X_1, X_3) }{p(R_1 = 1 \mid R_2=1, R_3=1, X_2, X_3 ) \times p(R_2 = 1 \mid R_1=1, R_3=1, X_1, X_3) } \times \theta  \\ 
		&\hspace{1.5cm} - R_3 \times (1-R_1) \times  (1 - R_2) \Big] = 0.  
	\end{align*}
}%

\begin{figure}[t]
	\centering
	\begin{subfigure}{.5\textwidth}
		\centering
		\includegraphics[height=12cm, width=9cm]{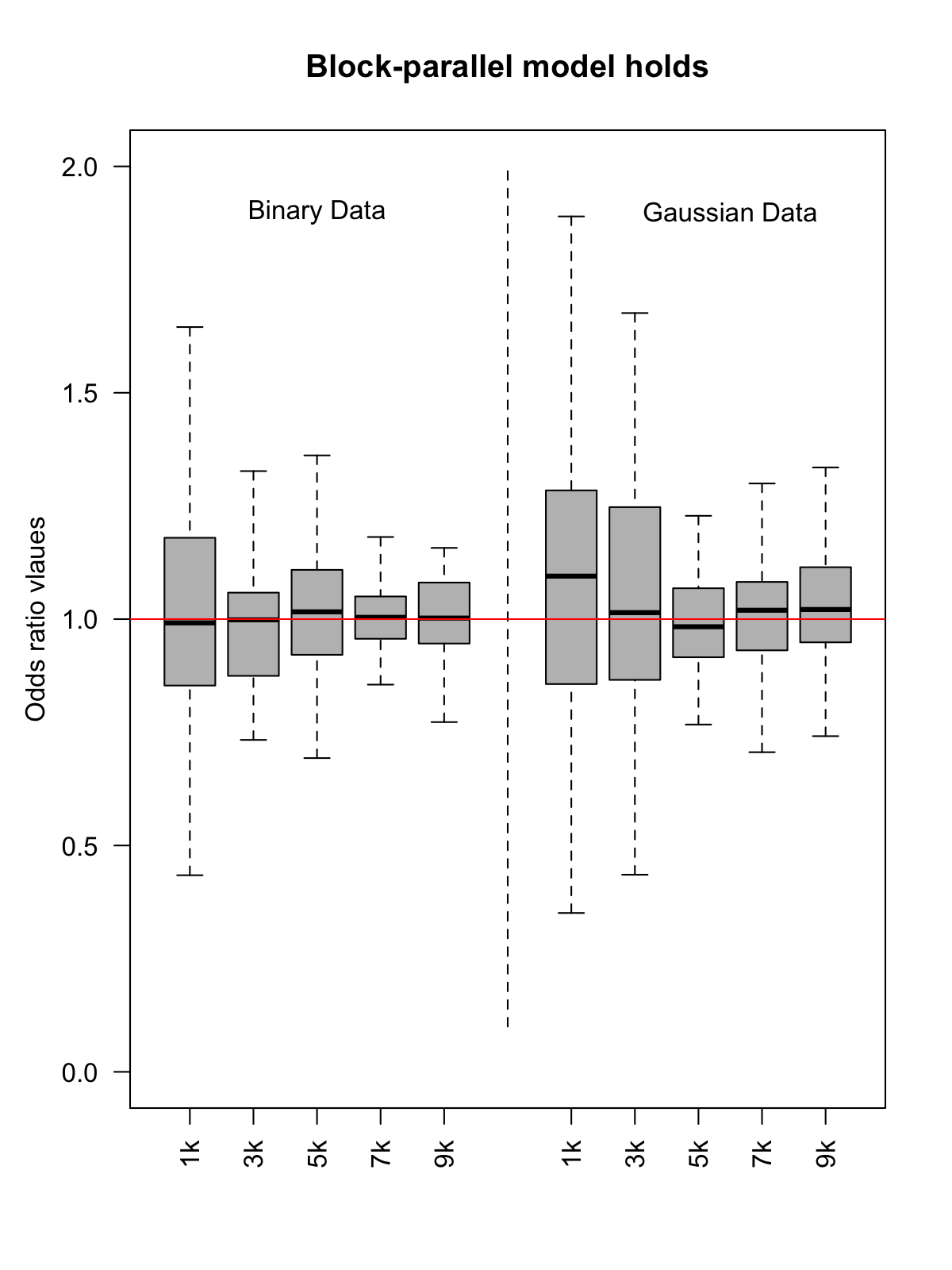}
		\label{fig:bp}
	\end{subfigure}%
	\begin{subfigure}{.5\textwidth}
		\centering
		\includegraphics[height=12cm, width=9cm]{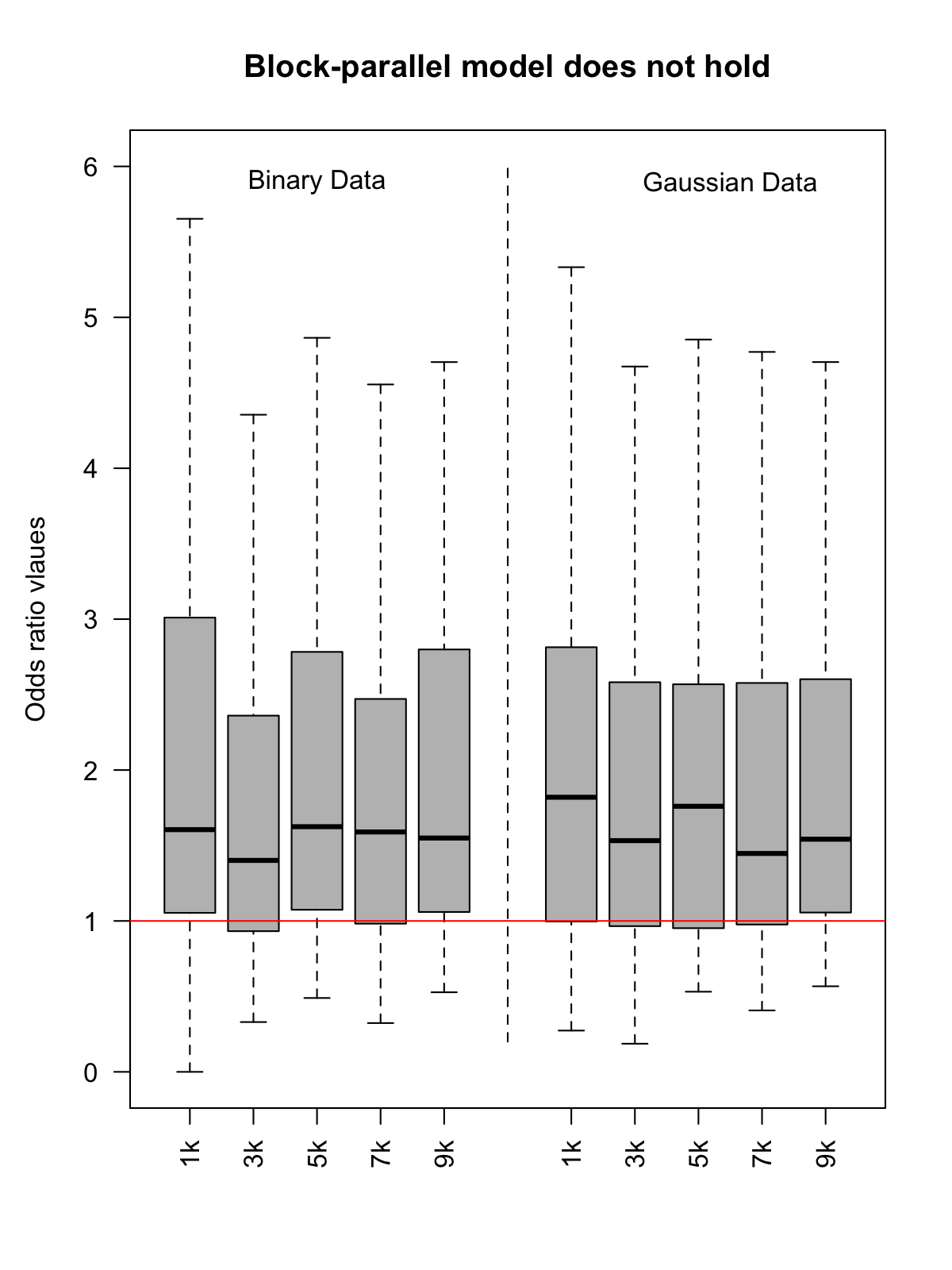}
		\label{fig:noself}
	\end{subfigure}
	\vspace{-1cm}
	\caption{Results on computing (conditional) odds ratio between a pair of missingness indicators to test an independence restriction between them. On the left panel, the block-parallel MNAR model captures the true underlying missingness mechanism. The assumptions of block-parallel MNAR model are violated on the right panel.} 
	\label{fig:bp_sim}
\end{figure}

For a fixed sample size, we simulate $100$ different datasets and calculate the odds ratio via the above estimating equation. We provide the boxplots  in Fig.~\ref{fig:bp_sim}. The x-axis is sample size that ranges from $1,000$ to $10,000$ with $2,000$ increments.  The left panel  illustrates the boxplots for binary and Gaussian data when the true missingness mechanism follows the restrictions of the block-parallel model, and in the right panel it does not. As it is shown, the boxplots are centered around $1$ in the left panel as expected, but move away from $1$ when the independence does not hold. To perform a formal test, we can construct confidence intervals for each sample size via bootstrapping the data generations and odds ratio calculations.

\end{document}